\documentclass[11pt,a4paper]{article}
\pdfoutput=1

\usepackage{jheppub}
\usepackage{amsmath}
\usepackage{amssymb}
\usepackage{graphics}
\usepackage[active]{srcltx}
\usepackage{pdfsync}
\usepackage{shuffle}
\usepackage{slashed}
\usepackage{hyperref}
\usepackage{subfigure}

\newcommand{\bea}{\begin{eqnarray}}
\newcommand{\eea}{\end{eqnarray}}
\newcommand{\nn}{\nonumber \\}

\def\W #1{\widetilde{#1}}

\def\eref#1{(\ref{#1})}

\def\a{{\alpha}}

\def\b{{\beta}}

\allowdisplaybreaks

\title{Multi-trace YMS amplitudes from soft behavior}
 \author[a,c]{Yi-Jian Du} \author[b]{Kang Zhou}

\affiliation[a]{Department of Physics, School of Physics and Technology,
Wuhan University, \\
No.299 Bayi Road, Wuhan 430072, China}

\affiliation[b]{Center for Gravitation and Cosmology, College of Physical Science and Technology, Yangzhou University,\\
No.180, Siwangting Road, Yangzhou, 225009, P.R. China.}

\affiliation[c]{Hubei Key Laboratory of Nuclear Solid Physics, School of Physics and Technology, Wuhan University,\\
No.299 Bayi Road, Wuhan 430072, China}

\emailAdd{yijian.du@whu.edu.cn} \emailAdd{zhoukang@yzu.edu.cn}

\date{\today}
\abstract{Tree level multi-trace Yang-Mills-scalar (YMS) amplitudes have been shown to satisfy a recursive expansion formula, which expresses any YMS amplitude by those with fewer gluons and/or scalar traces. In an earlier work, the single-trace expansion formula has been shown to be determined by the universality of soft behavior. This approach is nevertheless not extended to multi-trace case in a straightforward way. In this paper, we derive the expansion formula of tree-level multi-trace YMS amplitudes in a bottom-up way: we first determine the simplest amplitude, the double-trace pure scalar amplitude which involves two scalars in each trace. Then insert more scalars to one of the traces. Based on this amplitude, we further obtain the double-soft behavior when the trace containing only two scalars is soft. The multi-trace amplitudes with more scalars and more gluons finally follow from the double-soft behavior as well as the single-soft behaviors which has been derived before.
}


\begin{document}

\maketitle \flushbottom

\section{Introduction}
\label{sec-intro}

Scattering amplitudes were shown to satisfy the universal soft behavior, which states that an $n$-particle amplitude is factorized into an $(n-1)$-particle amplitude and a soft factor, when the momentum of an external particle tends to zero.  Such factorization behaviors were firstly proposed for photons and gravitons \cite{Low:1958sn,Weinberg:1965nx} via Feyman diagrams. By the help of the modern methods and formulations beyond Feynman diagrams such as Britto-Cachazo-Feng-Witten (BCFW) \cite{Britto:2004ap,Britto:2005fq} recursion relation and Cachazo-He-Yuan (CHY) formula \cite{Cachazo:2013gna,Cachazo:2013hca, Cachazo:2013iea, Cachazo:2014nsa,Cachazo:2014xea}, the soft behaviors for gravity (GR) \cite{Cachazo:2014fwa} and Yang-Mills (YM) theory \cite{Casali:2014xpa} were revived and further extended to higher-orders \cite{Cachazo:2014fwa,Schwab:2014xua,Afkhami-Jeddi:2014fia}. These progresses on soft theorems were subsequently generalized to string theory and to loop levels \cite{Bern:2014oka,He:2014bga,Cachazo:2014dia,Bianchi:2014gla,Sen:2017nim}.

An interesting observation on soft behavior is that one can inverse the soft limit and reconstruct a full amplitude. This construction has distinct aspects. On one hand, the inverse soft limit (ISL) program \cite{Nguyen:2009jk,Arkani-Hamed:2009ljj,Boucher-Veronneau:2011rwd,Rodina:2018pcb,Ma:2022qja} which inherits the structure of BCFW recursion can be used to construct four-dimensional helicity amplitudes in YM, GR and EYM (Einstein-Yang-Mills) amplitudes. On another hand, amplitudes in effective theories arise from the Aler's zero \cite{Cheung:2014dqa,Luo:2015tat,Elvang:2018dco}, which is the leading soft behavior of effective theories. In a recent work \cite{Zhou:2022orv} by one author of the current paper, a new approach on the soft program has been proposed. In this approach, an amplitude is supposed to be decomposed into a combination of bi-adjoint scalar (BAS) amplitudes and the universality of soft behavior is assumed. With these two assumptions, one can determine the soft behavior and even the full amplitude. This approach has already reconstructed \cite{Zhou:2022orv} the expansion formulas of single-trace YMS, EYM, YM and GR amplitudes \cite{Stieberger:2016lng,Nandan:2016pya,Schlotterer:2016cxa,Fu:2017uzt,Chiodaroli:2017ngp,Teng:2017tbo}  successfully, and was applied to deriving the Bern-Carrasco-Johansson (BCJ) \cite{Bern:2008qj} relations at tree-level and one-loop level \cite{Wei:2023iay}.

Although, the construction of amplitudes from soft behavior along the line \cite{Zhou:2022orv} has been effectively applied in many situations, there is still a gap that obstacles the generalization to multi-trace YMS or EYM amplitudes \cite{Du:2017gnh} directly.  Particularly, if one tries to generate amplitudes with multiple traces using the method \cite{Zhou:2022orv}, it is not straightforward to determine the expansion basis. This is because the expansion basis is determined by soft limit recursively, but one cannot obtain a nontrivial soft limit of a scalar particle from a trace with only two scalars. In this work, we provide a bottom-up approach to the multi-trace YMS amplitudes. The main idea is following. We begin with the simplest multi-trace amplitude: double-trace YMS amplitude where each trace contains two scalars. This is generated from the $4$-point YM amplitude via dimensional reduction. We further use single-soft theorem to insert more scalars into one of the two traces, then determine the double-soft behavior for this double-trace pure scalar amplitude when the trace containing only two scalars become soft. Assuming the universality of this double-soft behavior, a YMS amplitude with an arbitrary number of traces, each of which contains only two scalars, is fully determined. Those YMS amplitudes involving more scalars in each trace and/or more gluons are finally fixed by single-soft behaviors of scalars and/or gluons, along the line of \cite{Zhou:2022orv}. Having the YMS construction, we can further determine the multi-trace EYM amplitudes through the well known double copy.




The structure of this paper is organized as follows. In section. \ref{sec-preparations}, we provide a brief introduction to the expansion formulas for single- and multi-trace amplitudes of YMS/EYM. Helpful conclusions of soft behaviors are also reviewed. The bottom-up construction is demonstrated in section. \ref{sec:BottomUp} via the single-trace YMS amplitudes. In section. \ref{sec-pures}, we focus on the expansion formula for multi-trace YMS amplitudes with pure scalar external lines. Using the double-soft behaviors, we demonstrate that the multi-trace pure scalar amplitudes can be expanded into a basis where a scalar trace is arranged into a Kleiss-Kuijf (KK) basis \cite{Kleiss:1988ne}. More external particles are further added using soft limit  and then a general multi-trace pure scalar amplitude is obtained. Based on the pure scalar amplitudes, we further derive amplitudes with external gluons in section. \ref{sec-YMS-sg}. A summary and more discussions are provided in section. \ref{sec-conclu}.

\section{A review of expansion formulas and soft behaviors}\label{sec-preparations}
In this section, we review the expansion formulas of single and multi-trace YMS amplitudes, as well as helpful results of soft behaviors of tree level BAS and single-trace YMS amplitudes. Various notations which will be used frequently in this paper are also introduced.

\subsection{The expansion formulas of single- and double-trace YMS amplitudes}

The YMS theory under consideration is the massless ${\rm YM}\oplus {\rm BAS}$ theory with Lagrangian \cite{Chiodaroli:2017ngp,Cachazo:2014xea}
\bea
{\cal L}_{{\rm YM}\oplus {\rm BAS}}&=&-{1\over4}\,F^{a}_{\mu\nu}\,F^{a\mu\nu}+{1\over2}\,D_\mu\phi^{Aa}\,D^{\mu}\phi^{Aa}
-{g^2\over4}\,f^{abe}\,f^{ecd}\,\phi^{Aa}\phi^{Bb}\phi^{Ac}\phi^{Bd}\nn
& &+{\lambda \over3!}\,F^{ABC}f^{abc}\,
\phi^{Aa}\phi^{Bb}\phi^{Cc}\,.~~\label{lag}
\eea
The indices $a$, $b$, $c$ and $d$ run over the adjoint representation of the gauge group. Scalar fields carry additional flavor indices $A$, $B$, $C$. The field strength and covariant derivative are defined in the usual way
\bea
& &F^{a}_{\mu\nu}=\partial_\mu\,A^a_\nu-\partial_\nu\,A^a_\mu+g\,f^{abc}\,A^b_\mu A^c_\nu\,,\nn
& &D_\mu\phi^{Aa}=\partial_\mu\phi^{Aa}+g\,f^{abc}\,A^b_\mu\phi^{Ac}\,.
\eea
The general tree amplitude of this theory includes both massless external scalars and massless external gluons.
Through the standard technic, one can decompose the gauge group factors to obtain
\bea
{\cal A}_{n+m}=\sum_{\sigma\in{\cal S}_{n+m}\setminus Z_{n+m}}\,{\rm Tr}[T^{a_{\sigma_1}}\cdots T^{a_{\sigma_{n+m}}}]\,{\cal A}_{n+m,1}(\sigma_1,\cdots,\sigma_{n+m})\,,~~\label{deco-color}
\eea
where ${\cal S}_{n+m}\setminus Z_{n+m}$ denotes un-cyclic permutations among $n$ external scalars and $m$ external gluons, $T^{a_{\sigma_i}}$ encode generators of the gauge group. Here ${\cal A}_{n+m}$ is the kinematic part of the amplitude without coupling constants. Meanwhile, decomposition of the flavor group factors leads to the structure of tree amplitudes which is similar to that of loop amplitudes in that, unlike pure gauge theories, it is not restricted to have only single-trace term:
\bea
{\cal A}_{n+m}&=&\sum_{\sigma\in{\cal S}_n\setminus Z_n}\,{\rm Tr}[T^{A_{\sigma_1}}\cdots T^{A_{\sigma_m}}]\,{\cal A}_{n+m,1}(\sigma_1,\cdots,\sigma_n)\nn
& &+\sum_{n_1+n_2=n}\,\sum_{\a\in{\cal S}_{n_1}\setminus Z_{n_1}}\,\sum_{\b\in{\cal S}_{n_2}\setminus Z_{n_2}}\,{\rm Tr}[T^{A_{\a_1}}\cdots T^{A_{\a_{n_1}}}]\,{\rm Tr}[T^{A_{\b_1}}\cdots T^{A_{\b_{n_2}}}]\,{\cal A}_{n+m,2}(\a_1,\cdots,\a_{n_1}|\b_1,\cdots,\b_{n_2})\nn
& &+\cdots\,,~~\label{multi-trace}
\eea
where ${\cal S}_{n_i}\setminus Z_{n_i}$ again stands for the set of un-cyclic permutations. The single-trace and double-trace terms are collected on the first and second lines respectively, and $\cdots$ on the third line denotes the remaining multi-trace terms. External scalars for the partial amplitudes ${\cal A}_{n+m,2}(\a_1,\cdots,\a_{n_1}|\b_1,\cdots,\b_{n_2})$ on the second line are grouped into two orderings which are $\a_1,\cdots,\a_{n_1}$
and $\b_1,\cdots,\b_{n_2}$. Analogously, partial amplitudes in multi-trace terms on the third line include more orderings among external scalars.
From now on, we denote such tree amplitudes as ${\cal A}_{\rm YMS}(\pmb{1}|\cdots|\pmb{m};\{g_k\}||\pmb{\sigma})$, where $\pmb\sigma$ stands for the ordering among all external legs obtained by decomposing the gauge group factors, $\pmb{1},\cdots,\pmb{m}$ are ordered sets associated to scalar traces arise from the flavor group, and $\{g_k\}$ with $k\in\{1,\cdots,h\}$ denotes the unordered set of $h$ external gluons. Sometimes, we will write down a trace explicitly. For example, for the single-trace amplitude
${\cal A}_{\rm YMS}(\pmb{1};\{g_k\}||\pmb\sigma)$ with $\pmb{1}=\{1,\cdots,n\}$, we frequently use ${\cal A}_{\rm YMS}(1,\cdots,n;\{g_k\}||\pmb\sigma)$.

Tree-level single-trace YMS amplitudes ${\cal A}_{\rm YMS}\left(1,\cdots,n;\{g_k\}||\pmb{\sigma}\right)$ with the scalar trace $\{1,\cdots,n\}$ and the gluon set $\{g_k\}$ satisfy the following recursive expansion formula
\bea
&&{\cal A}_{\rm YMS}(1,\cdots,n;\{g_k\}||\pmb{\sigma})\nn
&=&\sum\limits_{\shuffle}\sum\limits_{i_1,...,i_l\in \{g_k\}\setminus p}\left(\epsilon_p\cdot f_{i_1}\cdot ...\cdot f_{i_l}\cdot Y_{i_l}\right)\,{\cal A}_{\rm YMS}(1,\{2,...,n-1\}\shuffle\{i_l,...,i_1,p\},n;\{g_k\}\setminus\{i_l,...,i_1,p\}||\pmb{\sigma}), \label{Eq:SingleExp}
\eea
where we have chosen an arbitrary element $p\in \{g_k\}$ as the fiducial particle, and the summation $\sum\limits_{\{i_1,...,i_l\}\in \{g_k\}\setminus p}$ means that we sum over all possible subsets of $\{g_k\}\setminus p$ and all possible permutations of elements in this subset. The summation $\sum\limits_{\shuffle}$ means we sum over all permutations such that the relative order in each of the ordered set $A$, $B$, $C$... is kept if the ordering of scalars in the amplitudes has the form $A\shuffle B\shuffle C\shuffle\dots$. In the coming discussions, this summation is always hidden and is implied by the shuffling ordering in the amplitude for short. The $\epsilon^{\mu}_p$ and $f_{i}^{\mu\nu}$  denote the polarization of the gluon $p$ and the strength tensor $f_{i}^{\mu\nu}\equiv k_i^{\mu}\epsilon_i^{\nu}-k_i^{\nu}\epsilon_i^{\mu}$, respectively.

The expansion formula of tree-level multi-trace YMS amplitude with pure scalar external particles ${\cal A}_{\rm YMS}(\pmb{1}|...|\pmb{m}||\pmb{\sigma})$ is given by
\bea
{\cal A}_{\rm YMS}(\pmb{1}|...|\pmb{m}||\pmb{\sigma})&=&\sum_{\substack{\pmb{i_1},...,\pmb{i_s}\\ \in\pmb{\text{Tr}}\setminus\{\pmb{1}, \pmb{2}\}} }\,\underset{\substack{c_j,d_j\in\pmb{i}_j\\ \text{for $j=1,...,s$}}}{\widetilde{\sum}}\,C_{c,d}(\pmb{K})\,{\cal A}_{\rm YMS}\big(1,\{2,...,n-1\}\shuffle\{c_s, \mathsf{KK}[\pmb{i}_s,c_s,d_s], d_s,\label{Eq:MultiPureScalar}\\
&&~~~~~~~~~~~~~~~~~~~~~~~~~~~..., c_1, \mathsf{KK}[\pmb{i}_1,c_1,d_1], d_1, a,\mathsf{KK}[\pmb{2},a,b], b\},n|\pmb{u}_1|...|\pmb{u}_{m-s-2}||\pmb\sigma\big),\nonumber
\eea
in which the first summation is taken over all possible choices of scalar traces $\pmb{i_1},...,\pmb{i_s}$ (including different permutations of them and all possible $2<s<m$) from the set of all scalar traces. The second summation means that we sum over all choices of ordered pair $\{c_j, d_j\}$ from the trace $\pmb{i}_j$ and if for a given pair, the trace is written as
\bea
\pmb{i}_j=\{c_j,\pmb{\alpha},d_j,\pmb{\beta}\},
\eea
we sum over permutations  $\pmb{\alpha}\shuffle\pmb{\beta}^T$, with an extra sign $(-1)^{n_{\pmb{\beta}}}$. The amplitudes on the r.h.s are those with $m-s-2$ traces, where the traces $\pmb{i}_1,...,\pmb{i}_s$ as well as the trace $\pmb{2}$ have been merged into the trace $\pmb{1}$. Each $\mathsf{KK}[\pmb{i}_j,c_j,d_j]$ correspondingly denotes the permutations
\bea
 \pmb{\alpha}\shuffle\pmb{\beta}^T.
\eea
The expansion coefficient $C_{a,b}(\pmb{K})$ is defined by\footnote{A sign is different from that in \cite{Du:2017gnh}.}
\bea
C_{c,d}(\pmb{K})=k_a\cdot (k_{b_1} k_{a_1})\cdot (k_{b_2} k_{a_2})\cdot ...\cdot (k_{b_s} k_{a_s})\cdot Y_{a_s},
\eea
where $Y_{a_s}$ is the total momentum of the scalars on the left of the gluon $a_s$, in YMS amplitudes on the r.h.s of (\ref{Eq:MultiPureScalar}).

When external gluons are added, the expansion formula has the form which combines eq. \eqref{Eq:SingleExp} and eq. \eqref{Eq:MultiPureScalar} together:
\bea
{\cal A}_{\rm YMS}(\pmb{1}|\pmb{2}|\cdots|\pmb{m};\{g_k\}||\pmb\sigma)&=&\underset{\pmb{\rm g}\subset\{g_k\}}{\sum}\sum_{\pmb{\rm Tr}_s\subset\pmb{\rm Tr}\setminus\{\pmb{1},\pmb{2}\}}\underset{\substack{c_j,d_j\in\pmb{i}_j\\ \text{for}\,j=1,...,s}}{\widetilde{\sum}}\underset{\substack{a\in \pmb{2}\\ a\neq b}}{\widetilde{\sum}}C_{c,d}(\pmb{K}\W\shuffle\pmb{\rm g}){\rm A}\big(\pmb{\rm Tr}_s\W\shuffle\pmb{\rm g}, a,\mathsf{KK}[\pmb{2},a,b], b;\{g_k\}\setminus\pmb{\rm g}\big),~~~~\label{expan-hg}\nn
\eea
where  $\pmb{\rm g}$ and $\pmb{\rm Tr}_s$ are ordered sets whose elements belong to the gluon set $\{g_k\}$ and $\pmb{\rm Tr}\setminus\{\pmb{1},\pmb{2}\}$, respectively. The $\pmb{K}$ is defined by the following ordered set
\bea
\pmb{K}=\big\{c_s, \mathsf{KK}[\pmb{i}_s,c_s,d_s], d_s,..., c_1, \mathsf{KK}[\pmb{i}_1,c_1,d_1], d_1\big\}.
\eea
The explicit expression of expansion coefficient $C_{a,b}(\pmb{K}\W\shuffle\pmb{\rm g})$ is given by
\bea
C_{c,d}\,(\pmb{K}\W\shuffle\pmb{\rm g})=k_{a}\cdot T_{\rho_1}\cdot T_{\rho_2}\cdots T_{\rho_{s+h}}\cdot Y_{\rho_{s+h}}\,,~~~\label{defin-C-2}
\eea
where we defined $\W\shuffle$ for the shuffling permutations of ordered sets of scalar traces (each should be considered as a single object) and gluons, and  $\pmb{\rho}=\pmb{\rm{g}}\W\shuffle\pmb{\rm Tr}_s$. The tensors $T_{\rho_r}^{\mu\nu}$ are introduced as
\bea
T_{\rho_r}^{\mu\nu}=\begin{cases}\displaystyle ~f_g^{\mu\nu}~~~~ & \rho_r=g\in\pmb{\rm{g}} \,,\\
\displaystyle ~k_{d_j}^\mu k_{c_j}^\nu~~~~~ & \rho_r=\pmb{i}_j\in\pmb{\rm Tr}_s\,.\end{cases}
\eea
The ${\rm A}(\pmb{\rm Tr}_s\W\shuffle\pmb{\rm g}, a,\mathsf{KK}[\pmb{2},a,b], b;\pmb{G}\setminus\pmb{\rm g})$ is a proper combination of YMS amplitudes with fewer traces and/or gluons
\bea
&&{\rm A}(\pmb{\rm Tr}_s\W\shuffle\pmb{\rm g}, a,\mathsf{KK}[\pmb{2},a,b], b;\pmb{G}\setminus\pmb{\rm g})={\cal A}_{\rm YMS}(1,\{2,\cdots,n-1\}\shuffle\{\pmb{K}\W\shuffle \pmb{\rm g},a,\mathsf{KK}[\pmb{2},a,b], b\},n|\pmb{r}_1|\cdots|\pmb{r}_q||\pmb\sigma)\,.\label{defin-A-1}\nn
\eea

The expansions in this subsection were previously studied from various angles \cite{Fu:2017uzt,Teng:2017tbo,Du:2017kpo,Du:2017gnh,Feng:2019tvb,Zhou:2019mbe}. In subsequent sections, all these expansions will be reconstructed via a purely bottom up perspective.

\subsection{A review of the soft behaviors of BAS amplitudes and the single-trace YMS amplitudes }

The tree level bi-adjoint scalar (BAS) amplitudes ${\cal A}_{\rm BAS}(\pmb\sigma||\W{\pmb\sigma})$ exclusively feature propagators for massless scalars. Each amplitude is simultaneously planar with respect to two orderings $\pmb\sigma$ and $\W{\pmb\sigma}$. Each BAS amplitude carries an overall $\pm$ sign. In this paper, we adopt the convention that the overall sign is $+$ when two orderings are identical. This convention is different from that in \cite{Cachazo:2013iea}. The sign for amplitudes with two different amplitudes can be generated from the above reference one by counting flippings \cite{Cachazo:2013iea}.

The leading soft behavior of BAS amplitude ${\cal A}_{\rm BAS}(1,\cdots,n||\pmb\sigma)$ when $i$ is the soft particle with momentum $k_i^{\mu}\sim \tau k_i^{\mu}$ ($\tau\to 0$), is presented by
\bea
{\cal A}^{(0)_i}_{\rm BAS}(1,\cdots,n||\pmb\sigma)&=&S^{(0)_i}_s\,{\cal A}_{\rm BAS}(1,\cdots,i-1,\not{i},i+1,\cdots,n||\pmb\sigma\setminus i)\,.~~~\label{for-soft-fac-s}
\eea
In the above, the leading soft factor $S^{(0)_i}_s$ is
\bea
S^{(0)_i}_s={1\over \tau}\,\left(\,{\delta_{i(i+1)}\over s_{i(i+1)}}+{\delta_{(i-1)i}\over s_{(i-1)i}}\,\right)\,,~~~~\label{soft-fac-s-0}
\eea
where the operator $\delta_{ij}$ is determined by the relative orders of $i,j$ in the ordering $\pmb\sigma$. If $i$, $j$ are not adjacent to each other, $\delta_{ij}=0$. If $i$ and $j$ are two adjacent elements, we have $\delta_{ij}=1$ for $i\prec j$, $\delta_{ij}=-1$ for $j\prec i$. The notation $i\prec j$ for a given permutation including elements $i$, $j$ means the position of $i$ is on the left side of the position of $j$.

More generally, one can consider the tree BAS amplitude ${\cal A}_{\rm BAS}(\pmb\sigma||\W{\pmb\sigma})$ which carries two orderings $\pmb\sigma$ and $\W{\pmb\sigma}$. Defining $\delta_{ij}$ and $\W\delta_{ij}$ for orderings $\pmb\sigma$ and $\W{\pmb\sigma}$ respectively, the leading soft behavior can be reformulated as
\bea
{\cal A}^{(0)_i}_{\rm BAS}(\pmb\sigma||\W{\pmb\sigma})&=&S^{(0)_i}_s\,{\cal A}_{\rm BAS}(\pmb\sigma\setminus i||\W{\pmb\sigma}\setminus i)\,,~~~\label{soft-s1}
\eea
with the soft factor
\bea
S^{(0)_i}_s={1\over\tau}\,\sum_{j\neq i}\,{\delta_{ij}\,\W\delta_{ij}\over s_{ij}}\,.~~\label{soft-fac-s1}
\eea

The antisymmetry of the operator  $\delta_{ij}$  results in the following property
\bea
& &\left({\delta_{ab}\over s_{ab}}+{\delta_{be}\over s_{be}}\right)\,{\cal A}_{\rm BAS}(1,\cdots,a,k_1,\cdots,k_p,e,\cdots,n||\pmb\sigma\setminus b)\nn
&=&\left({\delta_{ab}\over s_{ab}}+{\delta_{bk_1}\over s_{bk_1}}\right)\,{\cal A}_{\rm BAS}(1,\cdots,a,k_1,\cdots,k_p,e,\cdots,n||\pmb\sigma\setminus b)\nn
& &+\sum_{i=1}^{p-1}\,\left({\delta_{k_ib}\over s_{k_ib}}+{\delta_{bk_{i+1}}\over s_{bk_{i+1}}}\right)\,{\cal A}_{\rm BAS}(1,\cdots,a,k_1,\cdots,k_p,e,\cdots,n||\pmb\sigma\setminus b)\nn
& &+\left({\delta_{k_pb}\over s_{k_pb}}+{\delta_{be}\over s_{be}}\right)\,{\cal A}_{\rm BAS}(1,\cdots,a,k_1,\cdots,k_p,e,\cdots,n||\pmb\sigma\setminus b)\,,~~\label{tech}
\eea
which will be applied frequently in the current paper.

Based on the definition of tree BAS amplitudes, the leading double-soft behavior when two legs $i$ and $j$ are taken to be soft simultaneously is also obvious,
\bea
{\cal A}^{(0)_{ij}}_{\rm BAS}(\pmb\sigma||\W{\pmb\sigma})&=&S^{(0)_{ij}}_s\,{\cal A}_{\rm BAS}(\pmb\sigma\setminus \{ij\}||\W{\pmb\sigma}\setminus \{ij\})\,,~~~\label{soft-2s1}
\eea
with the soft factor
\bea
S^{(0)_{ij}}_s={1\over\tau^3}\,\sum_{k\neq i,j}\,{\delta_{k(ij)}\,\delta_{ij}\,\W\delta_{k(ij)}\,\W\delta_{ij}\over s_{ij}\,s_{ijk}}\,,~~\label{soft-fac-2s}
\eea
arises from diagrams those $a$ and $b$ couple to a common vertex, then couple to another external leg.
Here the new symbol $\delta_{k(ij)}$ depends on the ordering $\pmb\sigma$ is introduced as follows. We regard two legs $\{i,j\}$ as a single element when $\delta_{ij}\neq0$. When $k$ is adjacent to $(ij)$ in the ordering {$\pmb\sigma$}, then $\delta_{k(ij)}=+1$ or $\delta_{k(ij)}=-1$ if $k\prec(ij)$ or $(ij)\prec k$ respectively. If $k$ is not adjacent to $(ij)$, $\delta_{k(ij)}=0$. The definition for $\W\delta_{k(ij)}$ depends on $\W{\pmb\sigma}$ is analogous.

When the gluon $p$ is the soft particle (i.e. $k^{\mu}_p\to \tau k^{\mu}_p$ ($\tau\to 0$)), the leading and subleading soft gluon behavior is
\bea
{\cal A}_n(\pmb\sigma)=\Bigl[S^{(0)_{p}}_g+S^{(1)_{p}}_g\Bigr]\,{\cal A}_{n-1}(\pmb\sigma\setminus p)\,,
\eea
where ${\cal A}$ is a general tree amplitude includes the external gluon $p$. For instance, for the
single-trace YMS amplitude ${\cal A}_{\rm YMS}(1,\cdots,n;\{g_k\}|\pmb\sigma)$, the soft behavior is presented by
\bea
{\cal A}_{\rm YMS}(1,\cdots,n;\{g_k\}||\pmb\sigma)&=&\Bigl[S^{(0)_{p}}_g+S^{(1)_{p}}_g\Bigr]\,{\cal A}_{\rm YMS}(1,\cdots,n;\{g_k\}\setminus p||\pmb\sigma\setminus p)+{\cal O}(\tau).~~~~\label{soft-impose}
\eea
In the above, the leading and subleading soft gluon behaviors are respectively given by
\bea
S^{(0)_{p}}_g&=&{1\over\tau}\,\sum_{a\in\{1,\cdots,n\}\cup\{g_k\}\setminus p}\,{\delta_{ap}\,(\epsilon_{p}\cdot k_a)\over s_{ap}}\,,\nn
S^{(1)_{p}}_g&=&\sum_{a\in\{1,\cdots,n\}\cup\{g_k\}\setminus p}\,{\delta_{ap}\,\big(\epsilon_{p}\cdot J_a\cdot k_{p}\big)\over s_{ap}}\,,~~~\label{soft-fac-g-0-2}
\eea
with $\delta_{ap}$ defined by the ordering $\pmb\sigma$.
Here, the angular momentum $J_a^{\mu\nu}$ acts  on $k^\rho_a$  and on $\epsilon^\rho_a$ with the orbital part and the spin part of the generator correspondingly, as follows

\bea
J_a^{\mu\nu}\,k_a^\rho=k_a^{[\mu}\,{\partial k_a^\rho\over\partial k_{a,\nu]}}\equiv k_a^{\mu}\,{\partial k_a^\rho\over\partial k_{a,\nu}}-k_a^{\nu}\,{\partial k_a^\rho\over\partial k_{a,\mu}}\,,~~~~
J_a^{\mu\nu}\,\epsilon_a^\rho=\big(\eta^{\nu\rho}\,\delta^\mu_\sigma-\eta^{\mu\rho}\,\delta^\nu_\sigma\big)\,\epsilon^\sigma_a\,.
\eea
Helpful properties of $S^{(1)_p}_g$ that will be frequently used in this paper are displayed as \cite{Zhou:2022orv}
\bea
\big(S^{(1)_p}_g\,V_1\big)\cdot V_2={\delta_{p1}\over s_{p1}}\,(V_1\cdot f_p\cdot V_2)\,,~~~~
V_1\cdot\big(S^{(1)_p}_g\,V_2\big)={\delta_{2p}\over s_{2p}}\,(V_1\cdot f_p\cdot V_2)\,,~~~~\label{iden-1}
\eea
and
\bea
V_1\cdot\big(S^{(1)_p}_g\,f_a\big)\cdot V_2={\delta_{ap}\over s_{ap}}\,V_1\cdot(f_p\cdot f_a-f_a\cdot f_p)\cdot V_2\,,~~~~\label{iden-2}
\eea
for two arbitrary Lorentz vectors $V_1$ and $V_2$.

In the story of this paper, the leading soft behavior of BAS scalar is determined by the definition of double color-ordered tree BAS amplitudes, while the soft theorems for gluon are derived in the bottom up construction, as will be seen in the next section.

\section{A bottom-up construction for single-trace YMS and pure YM amplitudes}\label{sec:BottomUp}
As a warm-up, we use a bottom-up perspective to study the single-trace YMS and YM. The logic is different from the construction in \cite{Zhou:2022orv,Wei:2023yfy,Hu:2023lso}, in which the expansion to BAS amplitudes is directly given as an assumption, and coefficients are assumed to depend on only one of two orderings carried by BAS amplitudes. These properties will emerge naturally in the construction in this section.

\subsection{YMS amplitudes with one external gluon}
\label{subsec-1gluon}

In this subsection, we review the bottom up construction for the single-trace YMS amplitudes which contain external scalars $i\in\{1,\cdots,n\}$ and only one external gluon $p$.

The starting point is the bootstrapping for $3$-point ones ${\cal A}_{\rm YMS}(1,2;p||\pmb\sigma)$.
We can impose three conditions, which are the correct mass dimension, the linear dependence on the polarization $\epsilon_p$ carried by the gluon $p$, as well as the absence of pole since the $3$-point amplitude can never factorize into lower-point ones. These three conditions uniquely fix ${\cal A}_{\rm YMS}(1,2;p||1,p,2)$ to be $\epsilon_p\cdot k_1$, up to an overall sign\footnote{When talking about $3$-point amplitudes, we allow the components of external momenta to take complex values, otherwise the momentum conservation and on-shell conditions can not be satisfied simultaneously. A well known example is the MHV and anti-MHV YM amplitudes in spinor-helicity representation in $4$-d space-time.}. Notice that $\epsilon_p\cdot k_2$ is equivalent to $\epsilon_p\cdot k_1$, due to the momentum conservation and the on-shell condition $\epsilon_p\cdot k_p=0$. Using the observation ${\cal A}_{\rm BAS}(1,p,2||1,p,2)=1$, we arrive at the following expansion
\bea
{\cal A}_{\rm YMS}(1,2;p||1,p,2)=(\epsilon_p\cdot k_1)\,{\cal A}_{\rm BAS}(1,p,2||1,p,2)\,.~~\label{sYMS-3p-1}
\eea
 The YMS amplitude ${\cal A}_{\rm YMS}(1,2;p||2,p,1)$ can be generated from ${\cal A}_{\rm YMS}(1,2;p||1,p,2)$ by swamping legs $1$ and $2$, the antisymmetry of the structure constant $f^{abc}$ indicates ${\cal A}_{\rm YMS}(1,2;p||2,p,1)=-{\cal A}_{\rm YMS}(1,2;p||1,p,2)$. Therefore,
\bea
{\cal A}_{\rm YMS}(1,2;p||2,p,1)=(\epsilon_p\cdot k_1)\,{\cal A}_{\rm BAS}(1,p,2||2,p,1)\,,~~\label{sYMS-3p-2}
\eea
since ${\cal A}_{\rm BAS}(1,p,2||2,p,1)=-{\cal A}_{\rm BAS}(1,p,2||1,p,2)$. The cyclic symmetry ensures that $1,p,2$ and $2,p,1$ are only different orderings for the $3$-point case, thus we can combine \eref{sYMS-3p-1} and \eref{sYMS-3p-2} to get
\bea
{\cal A}_{\rm YMS}(1,2;p||\pmb\sigma)=(\epsilon_p\cdot k_1)\,{\cal A}_{\rm BAS}(1,p,2||\pmb\sigma)\,,~~\label{start-point}
\eea

Now we invert the leading soft theorem for external scalars, to insert more scalars into ${\cal A}_{\rm YMS}(1,2;p||\pmb\sigma)$.
Such manipulation is based on the assumption of the universality of soft behavior. In other words, we assume that the soft theorem for the external scalar holds for any amplitude which includes the BAS scalar as the external particle, and the soft factor acts on other external particles in the unique manner. Back to the current case under consideration, we assume that the soft factor
\bea
S^{(0)_i}_s={1\over \tau}\,\sum_{j\neq i}\,{\delta_{ij}\,\W\delta_{ij}\over s_{ij}}
\eea
given in \eref{soft-fac-s1} holds for YMS amplitudes. The above soft factor includes $\delta_{ij}$ and $\W\delta_{ij}$, thus is associated to two orderings. Suppose each external gluon also belong to two orderings simultaneously, then the soft factor of BAS scalar also acts on external gluons, with the same $S^{(0)_i}_s$. Since gluons carry only one ordering, thus the soft factor of BAS scalar does not act on them. Based on the above explanation, we now
consider the $4$-point amplitude ${\cal A}_{\rm YMS}(1,2,3;p||\pmb\sigma)$ with external scalars $i\in\{1,2,3\}$ and external gluon $p$. The soft theorem leads to the following leading soft behavior when taking $k_2\to\tau k_2$ \cite{Wei:2023yfy},
\bea
{\cal A}_{\rm YMS}^{(0)_2}(1,2,3;p||\pmb\sigma)
&=&(\epsilon_p\cdot k_1)\,\left[{\cal A}^{(0)_2}_{\rm BAS}(1,2,p,3||\pmb\sigma)+{\cal A}^{(0)_2}_{\rm BAS}(1,p,2,3||\pmb\sigma)\right]\,.~~\label{4s-2soft}
\eea
This soft behavior implies that ${\cal A}_{\rm YMS}(1,2,3;p||\pmb\sigma)$ can be expanded as
\bea
{\cal A}_{\rm YMS}(1,2,3;p||\pmb\sigma)&=&C(1,2,p,3)\,{\cal A}_{\rm BAS}(1,2,p,3||\pmb\sigma)\nn
& &+C(1,p,2,3)\,{\cal A}_{\rm BAS}(1,p,2,3||\pmb\sigma)\,,
\eea
where coefficients $C(1,2,p,3)$ and $C(1,p,2,3)$ satisfy
\bea
C^{(0)_2}(1,2,p,3)=C^{(0)_2}(1,p,2,3)=\epsilon_p\cdot k_1\,.~~\label{constraint}
\eea
The constraint \eref{constraint} requires $C(1,2,p,3)$ and $C(1,p,2,3)$ to have the form
\bea
C(1,2,p,3)=\epsilon_p\cdot k_1+\a_1\,(\epsilon_p\cdot k_2)\,,~~~~C(1,p,2,3)=\epsilon_p\cdot k_1+\a_2\,(\epsilon_p\cdot k_2)\,.
\eea
Terms with higher order of $k_2$ are forbidden by the correct mass dimension. Notice that we restrict $C(1,2,p,3)$ and $C(1,p,2,3)$ to be polynomials without any pole, since ${\cal A}_{\rm BAS}(1,2,p,3||\pmb\sigma)$ and ${\cal A}_{\rm BAS}(1,p,2,3||\pmb\sigma)$ forms a Kleiss-Kuijf (KK) basis. As proved in in Appendix. \ref{sec-ap}, when expanding an amplitude with massless propagators to KK basis, one can always find a formula in which coefficients do not contain any pole.
To determine $\a_1$ and $\a_2$, one can consider the leading soft behavior of $k_1\to\tau k_1$ to obtain \cite{Wei:2023yfy}
\bea
{\cal A}^{(0)_1}_{\rm YMS}(1,2,3;p||\pmb\sigma)
&=&(\epsilon_p\cdot k_2)\,{\cal A}^{(0)_1}_{\rm BAS}(1,2,p,3||\pmb\sigma)\,.~~\label{4p-1soft}
\eea
The soft behavior in \eref{4p-1soft} indicates
\bea
C^{(0)_1}(1,2,p,3)=\epsilon_p\cdot k_2\,,~~~~C^{(0)_1}(1,p,2,3)=0\,.~~\label{constraint2}
\eea
Comparing \eref{constraint2} with \eref{constraint}, we find $\a_1=1$, $\a_2=0$, thus the full expansion of ${\cal A}_{\rm YS}(1,2,3;p||\pmb\sigma)$ is found to be
\bea
{\cal A}_{\rm YMS}(1,2,3;p||\pmb\sigma)&=&(\epsilon_{p}\cdot k_{12})\,{\cal A}_{\rm BAS}(1,2,p,3||\pmb\sigma)+(\epsilon_p\cdot k_1)\,{\cal A}_{\rm BAS}(1,p,2,3||\pmb\sigma)\nn
&=&(\epsilon_p\cdot Y_p)\,{\cal A}_{\rm BAS}(1,2\shuffle p,3||\pmb\sigma)\,.
\eea
where the combinatorial momentum $Y_p$ is defined as the summation over momenta carried by scalars at the l.h.s of $p$ in the color ordering $(1,2\shuffle p,3)$.

Repeating the above manipulation, one can determine the general
YMS amplitude ${\cal A}_{\rm YMS}(1,\cdots,n;p||\pmb\sigma)$ with an arbitrary number of external scalars and one external gluon $p$, in the expanded formula \cite{Wei:2023yfy}
\bea
{\cal A}_{\rm YMS}(1,\cdots,n;p||\pmb\sigma)=(\epsilon_p\cdot Y_p)\,{\cal A}_{\rm BAS}(1,\{2,\cdots,n-1\}\shuffle p,n||\pmb\sigma).~~~~\label{expan-1g-old}
\eea
As can be seen, the YMS amplitude is expanded to BAS amplitudes, with legs $1$ and $n$ are fixed at two ends in the first ordering. Such BAS amplitudes are called KK BAS basis \cite{Kleiss:1988ne}. The coefficients $\epsilon_p\cdot Y_p$ in the expansion are independent of the second ordering $\pmb\sigma$ carried by BAS amplitudes.

\subsection{Soft factors of gluon and YMS amplitudes with more external gluons}
\label{subsec-moregluon}

From the single-trace YMS amplitudes given in \eref{expan-1g-old}, one can derive the single-soft theorem for external gluons by taking $k_p$ to be soft. The resulted soft factors at leading and subleading orders are given by \cite{Zhou:2022orv}
\bea
S^{(0)_p}_g={1\over\tau}\,\sum_{i=1}^n\,{\delta_{ip}\,(\epsilon_p\cdot k_i)\over s_{ip}}\,,~~\label{soft-fac-g-01}
\eea
and
\bea
S^{(1)_p}_g=-\sum_{i=1}^n\,{\delta_{ip}\over s_{ip}}\,k_i\cdot f_p\cdot {\partial\over\partial k_i}\,,~~\label{soft-fac-g-11}
\eea
with $f_p^{\mu\nu}\equiv k^\mu_p\epsilon^\nu_p-\epsilon^\mu_p k^\nu_p$. The subleading order soft factor in \eref{soft-fac-g-11} can also be reformulated as
\bea
S^{(1)_p}_g=\sum_{i=1}^n\,{\delta_{ip}\,\big(\epsilon_p\cdot J_i\cdot k_p\big)\over s_{ip}}\,,~~\label{soft-fac-g-12}
\eea
where $J_i^{\mu\nu}$ stands for the angular momentum carried by external particles $i$.
For external scalars, two formulas \eref{soft-fac-g-11} and \eref{soft-fac-g-12} are equivalent to each other, since the angular momentum of a scalar does not contain the spin part.

We will invert the subleading soft theorem for the gluon to construct single-trace YMS amplitudes with more external gluons. The subleading soft factor of gluon contains $\delta_{ip}$ determined by the ordering $\pmb\sigma$, and $\pmb\sigma$ includes both scalars and gluons. Therefore, the assumption of universality indicates that this soft operator acts on external scalars and gluons in the same way. Thus, for the amplitude ${\cal A}_{\rm YMS}(1,\cdots,n;p,q||\pmb\sigma)$ with two external gluons encoded by $p$ and $q$, suppose we take $k_q$ to be soft, to describe the corresponding soft behavior, the summation $\sum_{i\in\{1,\cdots,n\}}$ in \eref{soft-fac-g-01}, \eref{soft-fac-g-11} and \eref{soft-fac-g-12} should be modified as $\sum_{i\in\{1,\cdots,n\}\cup p}$.

Now an ambiguity arises. The angular momentum of the external gluon contains the spin part, this part breaks the equivalence between two candidates \eref{soft-fac-g-11} and \eref{soft-fac-g-12}, forces us to make a choice. The correct one is \eref{soft-fac-g-12}, as can be seen through the following argument. Based on the universality discussed previously, the soft operator $S^{(1)_q}_g$ acts on the gluon $p$. Then the explicit formulas in \eref{soft-fac-g-11} and \eref{soft-fac-g-12} lead to that both two candidates contain $\epsilon_q\cdot k_p$, thus the amplitude ${\cal A}_{\rm YMS}(1,\cdots,n;p,q||\pmb\sigma)$ also contains the Lorentz invariant $\epsilon_q\cdot k_p$. The amplitude ${\cal A}_{\rm YMS}(1,\cdots,n;p,q||\pmb\sigma)$ should also contain $\epsilon_p\cdot k_q$, due to the symmetry. It means
the subleading soft operator $S^{(1)_q}_g$ should act on $\epsilon_p$, since $\epsilon_p\cdot k_q$ is accompanied by $\tau$ under the rescaling $k_q\to\tau k_q$, and provides the contribution at $\tau^0$ order when combining with propagators $1/s_{aq}$ with $a\in\{1,\cdots,n\}\cup p$. The candidate \eref{soft-fac-g-12} acts on $\epsilon_p$ as
\bea
S^{(1)_q}_g\,\epsilon_p^\mu=-{\delta_{pq}\over s_{pq}}\,(\epsilon_p\cdot f_q)^\mu\,,
\eea
while the candidate \eref{soft-fac-g-11} annihilates $\epsilon_p$. Thus the choice \eref{soft-fac-g-11} is excluded.

According to the universality of soft behavior, the subleading soft factor in \eref{soft-fac-g-12} is now generalized as
\bea
S^{(1)_p}_g=\sum_{a\in\{1,\cdots,n\}\cup\{g_k\}\setminus p}\,{\delta_{ap}\,\big(\epsilon_{p}\cdot J_a\cdot k_{p}\big)\over s_{ap}}\,,~~\label{soft-fac-g-1a}
\eea
where $\{g_k\}$ is the set of all external gluons. Another useful equivalent formula is
\bea
S^{(1)_p}_g=-\sum_{V_a}\,{\delta_{ap}\over s_{ap}}\,V_a\cdot f_{p}\cdot {\partial\over\partial V_a}\,,~~\label{soft-fac-g-1b}
\eea
where the summation is over all Lorentz vectors $V_a$ carried by external scalars or gluons, without distinguishing momenta and polarizations.
The alternative formula \eref{soft-fac-g-1b} holds as long as the physical amplitudes are linear in each polarization.

Inverting the subleading soft factor of gluon in \eref{soft-fac-g-1a} and \eref{soft-fac-g-1b}, one can find \cite{Zhou:2022orv}
\bea
{\cal A}^{(1)_q}_{\rm YMS}(1,\cdots,n;p,q||\pmb\sigma)
&=&(\epsilon_p\cdot Y_p)\,{\cal A}^{(1)_q}_{\rm YMS}(1,\{2,\cdots,n-1\}\shuffle p,n;q||\pmb\sigma)\nn
& &+\tau\,(\epsilon_p\cdot f_q\cdot Y_q)\,{\cal A}^{(0)_q}_{\rm BAS}(1,\{2,\cdots,n-1\}\shuffle \{q,p\},n||\pmb\sigma)\,,~~\label{soft-sYMS-2g}
\eea
which indicates
\bea
{\cal A}_{\rm YMS}(1,\cdots,n;p,q||\pmb\sigma)
&=&(\epsilon_p\cdot Y_p)\,{\cal A}_{\rm YMS}(1,\{2,\cdots,n-1\}\shuffle p,n;q||\pmb\sigma)\nn
& &+(\epsilon_p\cdot f_q\cdot Y_q)\,{\cal A}_{\rm BAS}(1,\{2,\cdots,n-1\}\shuffle \{q,p\},n||\pmb\sigma)\,.~~\label{expan-sYMS-2g}
\eea
In the above construction, we employed the subleading soft behavior rather than the leading one. The reason is that the factor $\epsilon_p\cdot f_q\cdot Y_q$ in the second line of \eref{expan-sYMS-2g} is accompanied with $\tau$ under the rescaling $k_q\to\tau k_q$, thus the corresponding term can not be detected at the leading order. On the other hand, the correct mass dimension forbids the coefficients in the expansion to have ${\cal O}(\tau^2)$ contributions without containing any pole, thus all terms can be detected at the subleading order.

Applying the above method recursively, one can construct the general single-trace YMS amplitude, in the following expanded formula \cite{Zhou:2022orv}
\bea
& &{\cal A}_{\rm YMS}(1,\cdots,n;\{g_k\}||\pmb\sigma)\nn
&=&\sum_{\vec{\pmb{\a}}}\,\big(\epsilon_p\cdot F_{\vec{\pmb{\a}}}\cdot Y_{\vec{\pmb{\a}}}\big)\,{\cal A}_{\rm YMS}(1,\{2,\cdots,n-1\}\shuffle \{\vec{\pmb{\a}},p\},n;\{g_k\}\setminus\{p\cup\pmb{\a}\}||\pmb\sigma)\,,~~\label{expan-YMS-recur}
\eea
where $p$ is the fiducial gluon which can be chosen as any element in $\{g_k\}$, and $\pmb{\a}$ are subsets of $\{g_k\}\setminus p$ which is allowed to be empty. When $\pmb{\a}=\{g_k\}\setminus p$, the YMS amplitudes in the second line of \eref{expan-YMS-recur} are reduced to pure BAS ones. The ordered set $\vec{\pmb{\a}}$ is generated from $\pmb{\a}$
by giving an order among elements in $\pmb{\a}$. The tensor $F^{\mu\nu}_{\vec{\pmb{\a}}}$ is defined as
\bea
F^{\mu\nu}_{\vec{\pmb{\a}}}\equiv\big(f_{\a_\ell}\cdot f_{\a_{\ell-1}}\cdots f_{\a_2}\cdot f_{\a_1}\big)^{\mu\nu}\,,
\eea
for $\vec{\pmb{\a}}=\{\a_1,\cdots\a_\ell\}$. The combinatorial momentum $Y_{\vec{\pmb{\a}}}$ is the summation of momenta carried by external scalars at the l.h.s of $\a_1$ in the color ordering $(1,\{2,\cdots,n-1\}\shuffle \vec{\pmb{\a}},n)$, where $\a_1$ is the first element in the ordered set $\vec{\pmb{\a}}$. The summation in \eref{expan-YMS-recur} is over all inequivalent ordered sets $\vec{\pmb{\a}}$. As can be observed, the coefficients $\epsilon_p\cdot F_{\vec{\pmb{\a}}}\cdot Y_{\vec{\pmb{\a}}}$ in the expansion are independent of the ordering $\pmb\sigma$.

\subsection{YM amplitudes}
\label{subsec-YM}

The pure YM amplitudes can be construct via the similar procedure. We first bootstrap the $3$-point amplitude ${\cal A}_{\rm YM}(1,2,3)$ as
\bea
{\cal A}_{\rm YM}(1,2,3)=(k_1\cdot\epsilon_2)\,(\epsilon_3\cdot\epsilon_1)+(k_2\cdot\epsilon_3)\,(\epsilon_1\cdot\epsilon_2)
+(k_3\cdot\epsilon_1)\,(\epsilon_2\cdot\epsilon_3)\,,~~\label{3p-expli}
\eea
due to the mass dimension, the linear dependence on each polarization, and the absence of pole. This amplitude can be expanded as
\bea
{\cal A}_{\rm YM}(1,2,3)&=&\big(\epsilon_3\cdot\epsilon_1\big)\,{\cal A}_{\rm YMS}(1,3;2||1,2,3)+\big(\epsilon_3\cdot f_2\cdot\epsilon_1\big)\,{\cal A}_{\rm BAS}(1,2,3||1,2,3)\,.~~\label{3p-expan1}
\eea
Swamping legs $1$ and $3$ gives ${\cal A}_{\rm YM}(3,2,1)=-{\cal A}_{\rm YM}(1,2,3)$, therefore
\bea
{\cal A}_{\rm YM}(3,2,1)&=&\big(\epsilon_3\cdot\epsilon_1\big)\,{\cal A}_{\rm YMS}(1,3;2||3,2,1)+\big(\epsilon_3\cdot f_2\cdot\epsilon_1\big)\,{\cal A}_{\rm BAS}(1,2,3||3,2,1)\,.~~\label{3p-expan2}
\eea
Combining \eref{3p-expan1} and \eref{3p-expan2} yields
\bea
{\cal A}_{\rm YM}(\pmb\sigma)&=&\big(\epsilon_3\cdot\epsilon_1\big)\,{\cal A}_{\rm YMS}(1,3;2||\pmb\sigma)+\big(\epsilon_3\cdot f_2\cdot\epsilon_1\big)\,{\cal A}_{\rm BAS}(1,2,3||\pmb\sigma)\,.~~\label{3p-expan}
\eea
Then, by inverting subleading soft theorem of external gluon, one can find the expansion of general $n$-point YM amplitude to be \cite{Hu:2023lso}
\bea
{\cal A}_{\rm YM}(\pmb\sigma)=\sum_{\vec{\pmb{\a}}}\,\big(\epsilon_n\cdot F_{\vec{\pmb{\a}}}\cdot\epsilon_1\big)\,{\cal A}_{\rm YMS}(1,\vec{\pmb{\a}},n;\{2,\cdots,n-1\}\setminus\pmb{\a}||\pmb\sigma)\,,~~\label{expan-YM-form1}
\eea
where the notations $\pmb{\a}$, $\vec{\pmb{\a}}$ and $F_{\vec{\pmb{\a}}}$ are explained after \eref{expan-YMS-recur} in the previous subsection.
Again, the coefficients $\epsilon_n\cdot F_{\vec{\pmb{\a}}}\cdot\epsilon_1$ are independent of the ordering $\pmb\sigma$.

\subsection{Remarks}
\label{subsec-remark}

Some remarks are in order.
First, the whole construction for YMS and YM amplitudes is based on bootstrapping $3$-point amplitudes and the assumption of the universality of soft behaviors. The resulted amplitudes exhibit some elegant properties. They can be expanded to KK BAS basis\footnote{Using the expansion \eref{expan-YMS-recur} recursively, one can expand both YMS amplitudes in \eref{expan-YMS-recur} and YM amplitudes in \eref{expan-YM-form1} to such basis.}, and the coefficients depend on only one ordering carried by BAS amplitudes in such basis. However, these properties are not assumed at the beginning, they serve as the consequence of our construction. Indeed, the second property mentioned above is equivalent to the well known double copy structure. Notice that the the unitarity and locality are implicitly assumed in the soft behavior. On the other hand, the gauge invariance is not assumed, it also emerges naturally.

Secondly, the resulted formulas of amplitudes depend on the formulas of starting points in the recursive constructions. For example, supposing we use fundamental BCJ relation to rewrite the expanded formula of single-trace YMS amplitudes in \eref{expan-1g-old} as
\bea
{\cal A}_{\rm YMS}(1,\cdots,n;p||\pmb\sigma)
&=&{k_n\cdot f_p\cdot Y_p\over k_n\cdot k_p}\,{\cal A}_{\rm BAS}(1,\{2,\cdots,n-1\}\shuffle p,n||\pmb\sigma)\,,~~~~\label{expan-1g-gi}
\eea
the same method leads to another formula of general YMS amplitudes \cite{Wei:2023yfy}
\bea
& &{\cal A}_{\rm YMS}(1,\cdots,n;\{g_i\}_m||\pmb\sigma)\nn
&=&\sum_{\vec{\pmb{\a}}}\,{k_n\cdot F_{\vec{\pmb{\a}}}\cdot Y_{\vec{\pmb{\a}}}\over k_n\cdot k_{p_1\cdots p_m}}\,{\cal A}_{\rm YMS}(1,\{2,\cdots,n-1\}\shuffle \vec{\pmb{\a}},n;\{g_i\}_m\setminus\pmb{\a}||\pmb\sigma)\,,~~\label{expan-YMS-gi}
\eea
which manifests the gauge invariance for any polarization and the permutation symmetry among external gluons, while breaking the manifest locality.
As shown in \cite{Wei:2023yfy}, the above expansion in \eref{expan-YMS-gi} is equivalent to that found by Cheung and Mangan in \cite{Cheung:2021zvb}.
Another example is, if we rewrite the expanded $3$-point YM amplitudes in \eref{3p-expan} as
\bea
{\cal A}_{\rm YM}(\pmb\sigma)&=&-{{\rm tr}\,\big(f_3\cdot f_1\big)\over k_3\cdot k_1}\,{\cal A}_{\rm YMS}(1,3;2||\pmb\sigma)\nn
& &-{{\rm tr}\,\big(f_3\cdot f_2\cdot f_1\big)\over k_3\cdot k_1}\,{\cal A}_{\rm BAS}(1,2,3||\pmb\sigma)\,,~~\label{expan-3p-II}
\eea
the same method gives rise to \cite{Hu:2023lso}
\bea
{\cal A}_{\rm YM}(\pmb\sigma)=-\sum_{\vec{\pmb{\a}}}\,{{\rm tr}\,\big(f_n\cdot F_{\vec{\pmb{\a}}}\cdot f_1\big)\over k_n\cdot k_1}\,{\cal A}_{\rm YMS}(1,\vec{\pmb{\a}},n;\{2,\cdots,n-1\}\setminus\pmb{\a}||\pmb\sigma)\,,~~\label{expan-YM-form2}
\eea
which also manifests the gauge invariance for each polarization while breaking the explicit locality. It is straightforward to see the equivalence between two expansions in \eref{expan-YM-form1} and \eref{expan-YM-form2}. Let us denote expanded YM amplitudes in \eref{expan-YM-form1} and \eref{expan-YM-form2} as ${\cal A}^{\rm I}_{\rm YM}(\pmb\sigma)$ and ${\cal A}^{\rm II}_{\rm YM}(\pmb\sigma)$ respectively, then a little algebra yields
\bea
{\cal A}^{\rm II}_{\rm YM}(\pmb\sigma)&=&{\cal A}^{\rm I}_{\rm YM}(\pmb\sigma)-{k_n\cdot\epsilon_1\over k_n\cdot k_1}\,A_n-{\epsilon_n\cdot k_1\over k_n\cdot k_1}\,B_n+{\epsilon_n\cdot\epsilon_1\over k_n\cdot k_1}\,C_n\,,~~\label{ABC}
\eea
where
\bea
A_n&=&\sum_{\vec{\pmb{\a}}}\,\Big(\epsilon_n\cdot F_{\vec{\pmb{\a}}}\cdot k_1\Big)\,{\cal A}_{\rm YMS}(1,\vec{\pmb{\a}},n;\{2,\cdots,n-1\}\setminus\pmb{\a}||\pmb\sigma)\,,\nn
B_n&=&\sum_{\vec{\pmb{\a}}}\,\Big(k_n\cdot F_{\vec{\pmb{\a}}}\cdot \epsilon_1\Big)\,{\cal A}_{\rm YMS}(1,\vec{\pmb{\a}},n;\{2,\cdots,n-1\}\setminus\pmb{\a}||\pmb\sigma)\,,\nn
C_n&=&\sum_{\vec{\pmb{\a}}}\,\Big(k_n\cdot F_{\vec{\pmb{\a}}}\cdot k_1\Big)\,{\cal A}_{\rm YMS}(1,\vec{\pmb{\a}},n;\{2,\cdots,n-1\}\setminus\pmb{\a}||\pmb\sigma)\,,~~\label{gauge-inva}
\eea
On the other hand, imposing the gauge invariance for gluon $1$ or $n$ in \eref{expan-YM-form1} yields
\bea
A_n=B_n=C_n=0\,,
\eea
thus we conclude ${\cal A}^{\rm II}_{\rm YM}(\pmb\sigma)={\cal A}^{\rm I}_{\rm YM}(\pmb\sigma)$.

Finally, an interesting observation is, our method always insert the external gluon in a manifestly gauge invariant manner, due to the property \eref{soft-fac-g-1b}. Thus, supposing we express the starting point before adding gluons in an explicitly gauge invariant formula, such as in \eref{expan-1g-gi} and \eref{expan-3p-II}, the resulted general formula will manifest the gauge invariance for each polarization, as can be seen in \eref{expan-YMS-gi} and \eref{expan-YM-form2}. Thus our method is an useful tool for constructing gauge invariant representations.

\section{Pure scalar multi-trace YMS amplitudes}
\label{sec-pures}
Now we are ready for constructing multi-trace amplitudes with pure external scalar particles. In this section, we first determine four-scalar double-trace amplitudes. Then insert more scalars to one of the trace by the help of single-soft theorem. To go further, we study the double-soft theorem, and then apply it to adding more scalar traces, each of which contains two scalars. Finally, by the help of single-soft behavior, we obtain the general expansion formula for multi-trace pure scalar amplitudes.
\subsection{The $4$-point double-trace amplitude}
\label{subsec-4p-2tr}

In this subsection we consider the simplest multi-trace YMS amplitude, the $4$-point double-trace amplitude ${\cal A}_{\rm YMS}(\pmb{1}|\pmb{2}||\pmb\sigma)$, where two sets associated with two scalar traces are given by $\pmb{1}=\{1,2\}$, $\pmb{2}=\{a,b\}$\footnote{The orderings of two sets $\pmb{1}$ and $\pmb{2}$ are meaningless due to the cyclic symmetry, thus it is not necessary to regard them as two ordered sets.}.
These amplitude serves as the starting point of our recursive construction.

The physically acceptable $4$-point double-trace amplitude with mass dimension $0$ is not unique, due to the freedom for choosing $4$-point interaction. Thus we need to choose a special one among a wide range of candidates. To completely determine ${\cal A}_{\rm YMS}(\pmb{1}|\pmb{2}||\pmb\sigma)$,
we first observe that the $3$-point single-trace YMS amplitude with one external gluon and two external scalars can be generated from the $3$-point YM one by performing the dimensional reduction. More explicitly, one can set $k_1$, $k_2$, $k_3$ and $\epsilon_3$ to be the $d$-dimensional vectors, while the only non-zero components of $\epsilon_1$ and $\epsilon_2$ lie in the $(d+1)^{\rm th}$ dimension, with $\epsilon_1\cdot\epsilon_2=1$. From the $d$-dimensional perspective, the $3$-point YM amplitude in \eref{3p-expli} with such special configuration of kinematics behaves as $\epsilon_3\cdot k_2$, coincide with the $3$-point single-trace YMS amplitude found by bootstrapping in subsection. \ref{subsec-1gluon} of section. \ref{sec:BottomUp}. This simple example shows that one extra dimension creates one length-$2$ trace, then a direct generalization is, two extra dimensions creates two length-$2$ traces. Thus, it is natural to require that the $4$-point double-trace amplitude, which serves as the lowest-point one in the current multi-trace case, is generated from the $4$-point YM amplitude via the similar dimensional reduction procedure. Using the result in \eref{expan-YM-form1}, the $4$-point YM amplitude is given as
\bea
{\cal A}_{\rm YM}(\pmb\sigma)&=&(\epsilon_b\cdot\epsilon_2)\,{\cal A}_{\rm YMS}(2,b;1,a||\pmb\sigma)\nn
&=&(\epsilon_b\cdot f_1\cdot\epsilon_2)\,{\cal A}_{\rm YMS}(2,1,b;a||\pmb\sigma)+(\epsilon_b\cdot f_a\cdot\epsilon_2)\,{\cal A}_{\rm YMS}(2,a,b;1||\pmb\sigma)\nn
&=&(\epsilon_b\cdot f_1\cdot f_a\cdot\epsilon_2)\,{\cal A}_{\rm BAS}(2,a,1,b||\pmb\sigma)+(\epsilon_b\cdot f_a\cdot f_1\cdot\epsilon_2)\,{\cal A}_{\rm BAS}(2,1,a,b||\pmb\sigma)\,.~~\label{YM-4p}
\eea
Now let the momenta of external legs to be $d$-dimensional vectors, the non-zero components of $\epsilon_a$ and $\epsilon_b$ lie in the $(d+1)^{\rm th}$ dimension, and the non-zero components of $\epsilon_1$ and $\epsilon_2$ lie in the $(d+2)^{\rm th}$ dimension, with $\epsilon_a\cdot\epsilon_b=\epsilon_1\cdot\epsilon_2=1$. Then the formula in \eref{YM-4p} is reduced to
\bea
{\cal A}_{\rm YMS}(\pmb{1}|\pmb{2}||\pmb\sigma)&=&(k_a\cdot k_1)\,{\cal A}_{\rm BAS}(1,a,b,2||\pmb\sigma)\,,~~\label{2tr-4p}
\eea
up to an overall $-$.

In the reminder of this section, our task is to determine multi-trace amplitudes with more than two traces and more scalars in each trace. One may try to do this by using the single-soft behavior for BAS scalar provided in \eref{soft-s1} and \eref{soft-fac-s1}, as what has been done in the single-trace case in section. \ref{sec:BottomUp}. This method can be used to insert more scalars into each trace. For example, one can insert scalars into $\pmb{1}$, with $\delta_{ij}$ and $\W\delta_{ij}$ in \eref{soft-fac-s1} correspond to orderings $\pmb{1}$ and $\pmb\sigma$ respectively. Nevertheless, this approach fails to insert new traces, since each trace requires at lest two scalars. Instead of single-soft behavior, we use the double-soft behavior to add more scalar traces. However, the double-soft behavior in \eref{soft-2s1} and \eref{soft-fac-2s} does not make sense, due to the difference between $4$-point double-trace YMS amplitude and $4$-point pure BAS one in \eref{2tr-4p}, which indicates the different interactions. Thus a new type of double-soft behavior for the length-$2$ trace is required. To achieve this goal, we first add more scalars to the trace $\pmb{1}$ according to the single-soft behavior, thus obtain a double-trace amplitude whose one trace contains more than two particles. The resulted amplitude allows one to extract the double-soft factor corresponding to the trace $\pmb{2}$. We then study the particular form of the double-soft behavior via this double-trace amplitude. Using the double-soft behavior, we further determine the expansion formula for amplitude with more scalar traces, all but one of them contains only two scalars. A general multi-trace amplitude is finally constructed by adding more scalars to those traces which contain only two particles, through single-soft behavior.

\subsection{Double-trace amplitudes with one length-$2$ trace}
\label{subsec-2trace-1more}

In this subsection, we construct double-trace amplitude ${\cal A}_{\rm YMS}(\pmb{1}|\pmb{2}||\pmb\sigma)$ where the trace $\pmb{1}$ includes more than two elements, by applying the leading order single-soft factor for BAS scalar.

Consider ${\cal A}_{\rm YMS}(\pmb{1}|\pmb{2}||\pmb\sigma)$ with $\pmb{1}=\{1,2,3\}$ and $\pmb{2}=\{a,b\}$. The leading-order behavior for
BAS scalar when the soft particle is $2$ is presented by
\bea
{\cal A}^{(0)_2}_{\rm YMS}(1,2,3|\pmb{2}||\pmb\sigma)&=&\left({\delta_{12}\over s_{12}}+{\delta_{23}\over s_{23}}\right)\,{\cal A}_{\rm YMS}(1,3|\pmb{2}||\pmb\sigma\setminus2)\nn
&=&(k_a\cdot k_1)\,\left({\delta_{12}\over s_{12}}+{\delta_{23}\over s_{23}}\right)\,{\cal A}_{\rm BAS}(1,a,b,3||\pmb\sigma\setminus2)\nn
&=&(k_a\cdot k_1)\,{\cal A}^{(0)_2}_{\rm BAS}(1,\{a,b\}\shuffle2,3||\pmb\sigma)\,,~~\label{2tr-5p-2soft}
\eea
where we have used (\ref{2tr-4p}) on the second line, the leading soft behavior of BAS amplitudes and the property (\ref{tech}) on the third line. Since our purpose is to seek ${\cal A}^{(0)_2}_{\rm YMS}(1,2,3|\pmb{2}||\pmb\sigma)$ in the expansion formula, we want to interpret the last line of \eref{2tr-5p-2soft} as the soft behaviors of terms in the expansion. In other words, we have
\bea
{\cal A}_{\rm YMS}(1,2,3|\pmb{2}||\pmb\sigma)=C(\shuffle)\,{\cal A}_{\rm BAS}(1,\{a,b\}\shuffle2,3||\pmb\sigma)\,.~~\label{2tr-5p-expan1}
\eea
The coefficients $C(\shuffle)$ satisfies $C^{(0)_2}(\shuffle)=k_a\cdot k_1$, thus take the form $C(\shuffle)=k_a\cdot k_1+{\cal O}(k_2)$ where ${\cal O}(k_2)$ is the Lorentz invariant which is linear in $k_2$. Notice that the mass dimension forbids $C(\shuffle)$ to have terms with higher order of $k_2$, without including any pole. To fix ${\cal O}(k_2)$, we further consider the soft behavior of the scalar $1$
\bea
{\cal A}^{(0)_1}_{\rm YMS}(1,2,3|\pmb{2}||\pmb\sigma)&=&\Big({\delta_{31}\over s_{31}}+{\delta_{12}\over s_{12}}\Big)\,{\cal A}_{\rm YMS}(2,3|\pmb{2}||\pmb\sigma\setminus1)\nn
&=&(k_a\cdot k_2)\,\Big({\delta_{31}\over s_{31}}+{\delta_{12}\over s_{12}}\Big)\,{\cal A}_{\rm BAS}(2,a,b,3||\pmb\sigma\setminus1)\nn
&=&-(k_a\cdot k_2)\,{\cal A}^{(0)_1}_{\rm BAS}(2,\{a,b\}\shuffle1,3||\pmb\sigma)\nn
&=&(k_a\cdot k_2)\,{\cal A}^{(0)_1}_{\rm BAS}(1,2,a,b,3||\pmb\sigma)\,,~~\label{2tr-5p-1soft}
\eea
where the last equality uses KK relation $\,{\cal A}_{\rm BAS}(1,\pmb{\alpha},n,\pmb{\beta}||\pmb\sigma)=(-1)^{n_{\pmb{\beta}}}{\cal A}_{\rm BAS}(1,\pmb{\alpha}\shuffle\pmb{\beta},n||\pmb\sigma)$, where $\pmb{\alpha}$ and $\pmb{\beta}$ are two ordered sets.
The soft behavior in \eref{2tr-5p-1soft} indicates
${\cal O}(k_2)=k_a\cdot k_2$ for the ordering $(1,2,a,b,3)$, and ${\cal O}(k_2)=0$ otherwise. Thus we find the expansion formula
\bea
{\cal A}_{\rm YMS}(1,2,3|\pmb{2}||\pmb\sigma)=(k_a\cdot Y_a)\,{\cal A}_{\rm BAS}(1,\{a,b\}\shuffle2,3||\pmb\sigma)\,,
\eea
which expresses the double-trace amplitude ${\cal A}_{\rm YMS}(1,2,3|\pmb{2}||\pmb\sigma)$ in terms of the single-trace ones where the scalars $1$ and $3$ play as the first and the last elements, respectively.

The above discussion can be extended to more general case. Supposing the $n+1$-point amplitude ${\cal A}_{\rm YMS}(\pmb{1}|\pmb{2}||\pmb\sigma)$, where $\pmb{1}$ is a length-$n-1$ set, can be expanded as
\bea
{\cal A}_{\rm YMS}(\pmb{1}|\pmb{2}||\pmb\sigma)=(k_a\cdot Y_a)\,{\cal A}_{\rm BAS}(1,\{a,b\}\shuffle\{2,\cdots,n-1\},n||\pmb\sigma)\,,~~\label{2tr-mp}
\eea
one can prove that such expansion formula also holds for $n+2$-point amplitudes. Consider the soft behavior of the scalar $2$. The leading order contribution of this soft behavior is
\bea
{\cal A}^{(0)_2}_{\rm YMS}(1,\cdots,n|\pmb{2}||\pmb\sigma)&=&\Big({\delta_{12}\over s_{12}}+{\delta_{23}\over s_{23}}\Big)\,{\cal A}_{\rm YMS}(1,3,\cdots,n|\pmb{2}||\pmb\sigma\setminus2)\nn
&=&(k_a\cdot Y_a)\,\Big({\delta_{12}\over s_{12}}+{\delta_{23}\over s_{23}}\Big)\,{\cal A}_{\rm BAS}(1,\{a,b\}\shuffle\{3,\cdots,n-1\},n||\pmb\sigma\setminus2)\nn
&=&(k_a\cdot Y^{(0)_2}_a)\,{\cal A}^{(0)_2}_{\rm BAS}(1,\{a,b\}\shuffle\{2,\cdots,n-1\},n||\pmb\sigma)\,,~~\label{2tr-m+3p-2soft}
\eea
where the last equality uses the anti-symmetry of $\delta_{ab}$ which indicates, for example,
\bea
{\delta_{12}\over s_{12}}+{\delta_{23}\over s_{23}}={\delta_{12}\over s_{12}}+{\delta_{2a}\over s_{2a}}+{\delta_{a2}\over s_{a2}}+{\delta_{23}\over s_{23}}\,.~~\label{example}
\eea
Notice that $Y_a$ on the second line of \eref{2tr-m+3p-2soft} are defined for orderings $1,\{a,b\}\shuffle\{3,\cdots,n-1\},n$, while on the third line are defined for
$1,\{a,b\}\shuffle\{2,3,\cdots,n-1\},n$. Apparently, $Y^{(0)_2}_a$ is generated from $Y_a$ by removing the component $k_2$.
The leading soft behavior in \eref{2tr-m+3p-2soft} indicates the following expansion formula
\bea
{\cal A}_{\rm YMS}(1,\cdots,n|\pmb{2}||\pmb\sigma)=C(\shuffle)\,{\cal A}_{\rm BAS}(1,\{a,b\}\shuffle\{2,\cdots,n-1\},n||\pmb\sigma)\,,
\eea
where the coefficients $C(\shuffle)$ satisfy
\bea
C^{(0)_2}(\shuffle)=k_a\cdot Y^{(0)_2}_a\,.~~\label{condi-1}
\eea
To fix $C(\shuffle)$, one can consider the leading soft behavior of another external particle, for example the scalar $3$, to obtain
\bea
C^{(0)_3}(\shuffle)=k_a\cdot Y^{(0)_3}_a\,.~~\label{condi-2}
\eea
Comparing \eref{condi-1} with \eref{condi-2}, we find the only solution is $C(\shuffle)=k_a\cdot Y_a$, thus the expansion formula \eref{2tr-mp}
is correct for the $n+2$-point case. Consequently, ${\cal A}_{\rm YMS}(\pmb{1}|\pmb{2}||\pmb\sigma)$ can be expressed by
\bea
{\cal A}_{\rm YMS}(\pmb{1}|\pmb{2}||\pmb\sigma)=(k_a\cdot Y_a)\,{\cal A}_{\rm BAS}(1,\{a,b\}\shuffle\{2,\cdots,n-1\},n||\pmb\sigma)\,,~~\label{2tr-general}
\eea
for $\pmb{1}=\{1,\cdots,n\}$ and $\pmb{2}=\{a,b\}$. Having this formula, we are ready for deriving the double-soft behavior corresponding to a length-2 trace containing the two soft scalars.

%
%

\subsection{Double-soft theorem}
\label{double-soft-theo}

Now we derive the double-soft theorem for $k_a\to\tau k_a,k_b\to\tau k_b$, where $a$, $b$ are elements of a length-2 trace, by using the expansion formula \eref{2tr-general}. We emphasize that all Feynman diagrams discussed in this subsection are diagrams for pure BAS amplitudes.

\subsubsection*{Leading order double-soft factor}

\begin{figure}
  \centering
  \includegraphics[width=6cm]{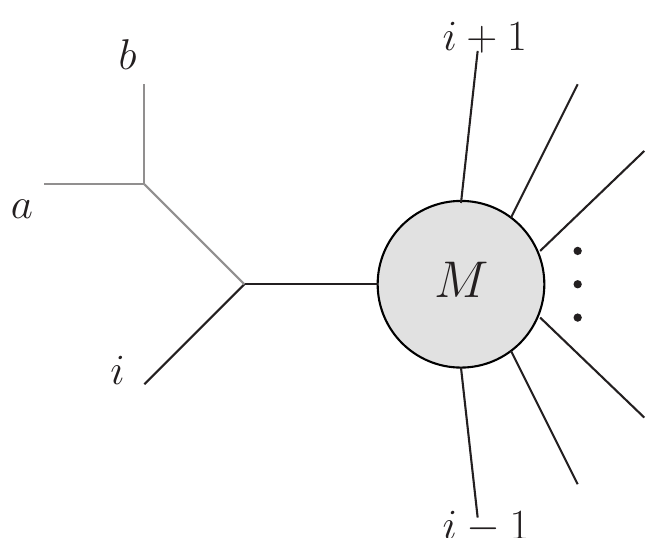} \\
  \caption{Feynman diagram contributes to the leading order double-soft behavior. The gray lines denote soft particles.}\label{leading}
\end{figure}

From \eref{2tr-general}, it is direct to observe that the leading order contribution arises from those Feynman diagrams where the external scalars $a$ and $b$ couple to the same vertex, and the propagator $1/s_{ab}$ couples to an external scalar $i$ of the pure BAS amplitude ${\cal A}_{\rm BAS}(1,\cdots,n||\pmb\sigma\setminus\pmb{2})$, as shown in Figure. \ref{leading}. This part reads
\bea
{\cal A}^{(0)_{ab}}_{\rm YMS}(\pmb{1}|\pmb{2}||\pmb\sigma)&=&\sum_{i=1}^{n-1}{1\over \tau^2}{\delta_{ab}\over s_{ab}}\left({\delta_{i(ab)}\over 2k_i\cdot k_{ab}}+{\delta_{(ab)(i+1)}\over 2k_{i+1}\cdot k_{ab}}\right)(k_a\cdot Y_a){\cal A}_{\rm BAS}(1,\cdots,i,\not{a},\not{b},i+1,\cdots,n||\pmb\sigma\setminus \pmb{2})\nn
&=&{1\over \tau^2}{\delta_{ab}\over s_{ab}}\sum_{i=1}^{n-1}\,\left[\,\sum_{j=i}^{n-1}\left({\delta_{j(ab)}\over 2k_j\cdot k_{ab}}+{\delta_{(ab)(j+1)}\over 2k_{j+1}\cdot k_{ab}}\right)\right](k_a\cdot k_i){\cal A}_{\rm BAS}(1,\cdots,n||\pmb\sigma\setminus \pmb{2})\nn
&=&{1\over \tau^2}\,{\delta_{ab}\over s_{ab}}\,\left[\,\sum_{i=1}^{n-1}\left({\delta_{i(ab)}\over 2k_i\cdot k_{ab}}+{\delta_{(ab)n}\over 2k_n\cdot k_{ab}}\right)(k_a\cdot k_i)\right]{\cal A}_{\rm BAS}(1,\cdots,n||\pmb\sigma\setminus \pmb{2})\nn
&=&{1\over \tau^2}\,{\delta_{ab}\over s_{ab}}\left[\,\sum_{i=1}^{n}\,{\delta_{i(ab)}\,(k_i\cdot k_a)\over 2k_i\cdot k_{ab}}\right]{\cal A}_{\rm BAS}(1,\cdots,n||\pmb\sigma\setminus \pmb{2}).~~~~\label{YS-ds-0}
\eea
Here the symbol $\delta_{i(ab)}$ is defined below \eref{soft-fac-2s}. According to the definition, we have
\bea
\delta_{i(ab)}=-\delta_{(ab)i}\,,~~~~\sum_{i=1}^n\,\delta_{i(ab)}=0\,.
\eea
The propagator $1/s_{ab}$ contributes $\tau^{-2}$, the propagator $1/(2k_i\cdot k_{ab})$ contributes $\tau^{-1}$, and the coefficient $k_a\cdot k_1$ contributes $\tau^{-1}$. Combining these contributions together leads to the overall factor $\tau^{-2}$.
In \eref{YS-ds-0},  the definition of $Y_a$ was used on the second line, the property $\delta_{i(ab)}=-\delta_{(ab)i}$ while $\delta_{ab}\neq0$ was applied on the third line, and  we applied $\delta_{ia}=-\delta_{bi}$ as well as momentum conservation on the fourth line. Notice that when using momentum conservation to get the last line, we ignored $k_a\cdot k_b$ since $k_b$ is accompanied by an extra factor $\tau$.
Thus we finally get the leading order double-soft theorem
\bea
{\cal A}^{(0)_{ab}}_{\rm YMS}(\pmb{1}|\pmb{2}||\pmb\sigma)=S^{(0)_{ab}}_s\,{\cal A}^{(0)_{ab}}_{\rm YMS}(\pmb{1}||\pmb\sigma\setminus\pmb{2})\,,
\eea
where the leading soft factor $S^{(0)_{ab}}_s$ is given by
\bea
S^{(0)_{ab}}_s={1\over \tau^3}{\delta_{ab}\over s_{ab}}\left[\,\sum_{i=1}^{n}\,{\delta_{i(ab)}\,(k_i\cdot k_a)\over 2k_i\cdot k_{ab}}\right]\,.
\eea

\subsubsection*{Subleading order double-soft factor}

We now turn to the subleading order $\tau^{-1}$ part of the double-soft behavior. To find the $\tau^{-1}$ term ${\cal A}^{(1)_{ab}}_{\rm YMS}(\pmb{1}|\pmb{2}||\pmb\sigma)$, we decompose the contributions into various pieces, and consider them one by one.

The first contribution also arises from the diagram in Figure. \ref{leading}. For a given scalar $i$, collecting the corresponding diagrams together yields
\bea
B_i(\tau)&=&{1\over \tau}{\delta_{ab}\over s_{ab}}{\,1\over 2\tau k_i\cdot k_{ab}+2\tau^2 k_a\cdot k_b}\,M_i(\tau)\nn
&=&{1\over \tau}{\delta_{ab}\over s_{ab}}{1\over 2\tau k_i\cdot k_{ab}}\left(M_i(0)+\tau{\partial\over \partial \tau}M_i(\tau)\Big|_{\tau=0}-\tau{k_a\cdot k_b\over k_i\cdot k_{ab}}M_i(0)+\cdots\right)\,,~~~~\label{B1}
\eea
where the block $M$ is denoted in Figure. \ref{leading}.
In the above expression, the overall scale factor $\tau^{-1}$ is counted as follows, the propagator $1/s_{ab}$ contributes $\tau^{-2}$, and the coefficient $\tau k_a\cdot Y_a$ in the expansion \eref{2tr-general} cancels one power. When expanding both numerator and denominator  according to powers of  $\tau$,  we get the second line. The first term on the second line of \eref{B1} describes the leading order single-soft behavior of a single-trace amplitude. Therefore
\bea
M_i(0)=\delta_{i(ab)}\,(k_i\cdot k_a)\,{\cal A}_{\rm BAS}(\pmb{1}||\pmb\sigma\setminus\pmb{2})\,,~~~~\label{M}
\eea
the scale parameter $\tau$ accompanied by $k_i\cdot k_a$ is absorbed into the overall $\tau^{-1}$ in \eref{B1}.
The $\tau^{-1}$ contributions of $B_i(\tau)$ are the second and the third terms on the second line at the r.h.s of \eref{B1} which arise from the expansions of $M_i(\tau)$ and the propagator $1/s_{iab}$, respectively. Now we focus on the second term first.
The parameter $\tau$ enters $M_i(\tau)$ only through the combination $k_i+\tau k_{ab}$, this observation indicates
\bea
{\partial\over \partial \tau}\,M_i(\tau)={1\over\tau}\,k_{ab}\cdot{\partial\over \partial (k_a+k_b)}\,M_i(\tau)=k_{ab}\cdot{\partial\over \partial k_i}\,M_i(\tau)\,.
\eea
Thus
\bea
{\partial\over \partial \tau}\,M_i(\tau)\Big|_{\tau=0}&=&k_{ab}\cdot{\partial\over \partial k_i}\,M_i(0)=\delta_{i(ab)}\,(k_i\cdot k_a)\,k_{ab}\cdot{\partial\over \partial k_i}\,{\cal A}_{\rm BAS}(\pmb{1}||\pmb\sigma\setminus\pmb{2})\,,~~~~\label{M'}
\eea
where we have applied \eref{M} to get the second equality. Substituting \eref{M'} into \eref{B1}, and summing over all external scalars $i$, we find the corresponding $\tau^{-1}$ order contribution
\bea
B_1={1\over\tau}\,{\delta_{ab}\over s_{ab}}\,\sum_{i=1}^n\,{\delta_{i(ab)}\,(k_i\cdot k_a)\over 2k_i\cdot k_{ab}}\,k_{ab}\cdot{\partial\over \partial k_i}\,{\cal A}_{\rm BAS}(\pmb{1}||\pmb\sigma\setminus\pmb{2})\,.~~~~\label{B-1}
\eea
The third term on the second line of \eref{B1} is directly written as
\bea
B_2=-{\delta_{ab}\over \tau}\,\left[\,\sum_{i=1}^n\,{\delta_{i(ab)}\,(k_i\cdot k_a)\over \big(2k_i\cdot k_{ab}\big)^2}\right]\,{\cal A}_{\rm BAS}(\pmb{1}||\pmb\sigma\setminus\pmb{2})\,.~~\label{B-2}
\eea
\begin{figure}
  \centering
  \includegraphics[width=8cm]{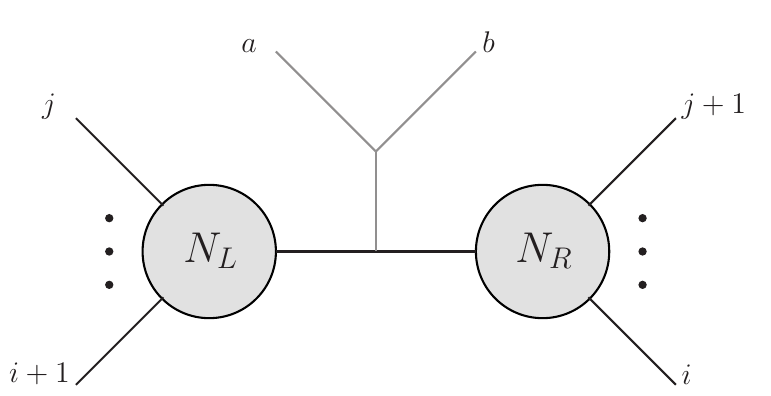} \\
  \caption{The second type of Feynman diagram which contributes to the subleading soft operator. The gray lines denote soft particles.}\label{NsN}
\end{figure}

The second case is that $a$ and $b$ couple to the same vertex and the propagator $1/s_{ab}$ couple to an internal propagator of the BAS amplitude ${\cal A}_{\rm BAS}(\pmb{1}||\pmb\sigma\setminus\pmb{2})$, see Figure.\ref{NsN}. Supposing $1/s_{ab}$ is coupled to the propagator $1/s_{(i+1)(i+2)\cdots(j-1)j}$ with $i<j$, only ${\cal A}_{\rm BAS}(1,\cdots,i,a,b,i+1,\cdots,n||\pmb\sigma)$ and ${\cal A}_{\rm BAS}(1,\cdots,j,a,b,j+1,\cdots,n||\sigma_{n+2})$ in the expansion \eref{2tr-general} carry the proper orderings.
Collecting contributions from these two amplitudes gives
\bea
D_{ij}(\tau)&=&{\rm sign}(\pm)\,{\delta_{ab}\over s_{ab}}\,\big(k_a\cdot K_{(i+1) j}\big)\,N_L(\tau)\,{1\over s_{(i+1)(i+2)\cdots(j-1)j}}\,{1\over s_{(i+1)(i+2)\cdots(j-1)jab}}\,N_R(\tau)\,,~~~~\label{NN}
\eea
where
\bea
s_{(i+1)(i+2)\cdots(j-1)jab}=s_{(i+1)(i+2)\cdots(j-1)j}+2\tau K_{(i+1)j}\cdot k_{ab}+2\tau^2 k_a\cdot k_b\,,
\eea
with $K_{(i+1)j}=k_{(i+1)(i+2)\cdots(j-1)j}$. Two building blocks $N_L$ and $N_R$ are shown in Figure.\ref{NsN}.
The factor $k_a\cdot K_{(i+1)j}$ arises as follows. Let us denote $Y_a$ for two cases ${\cal A}_{\rm BAS}(1,\cdots,j,a,b,j+1,\cdots,n||\pmb\sigma)$ and ${\cal A}_{\rm BAS}(1,\cdots,i,a,b,i+1,\cdots,n||\pmb\sigma)$
as $Y_{a;j}$ and $Y_{a;i}$, respectively. Since two configurations related to each other by swapping propagators $1/s_{ab}$ and
$1/s_{(i+1)(i+2)\cdots(j-1)j}$, the antisymmetry of structure constant $f^{abc}$ indicates a relative $-$ between two cases. Thus we
get
\bea
k_a\cdot (Y_{a;j}-Y_{a;i})=k_a\cdot K_{(i+1)j}\,.~~\label{Y12}
\eea
The sign denoted as ${\rm sign}(\pm)$ is obtained by decomposing the overall sign of the amplitude ${\cal A}_{\rm BAS}(1,\cdots,j,a,b,j+1,\cdots,n||\pmb\sigma)$ into ${\rm sign}(\pm)$ and the sign carried by $\delta_{ab}$ (Suppose we choose $Y_{a;i}-Y_{a;j}$ instead of $Y_{a;j}-Y_{a;i}$ in \eref{Y12}, then ${\rm sign}(\pm)$ should be determined by decomposing the overall sign of the amplitude ${\cal A}_{\rm BAS}(1,\cdots,i,a,b,i+1,\cdots,n||\pmb\sigma)$). In \eref{NN}, the $\tau^{-1}$ order contribution is
\bea
D^{(0)_{ab}}_{ij}&=&{\rm sign}(\pm)\,{\delta_{ab}\over \tau\,s_{ab}}\,{k_a\cdot K_{(i+1)j}\over s^2_{(i+1)(i+2)\cdots(j-1)j}}\,N_L(0)\,N_R(0)\nn
&=&-{\rm sign}'(\pm)\,{\delta_{ab}\over 4\tau\,s_{ab}}\,\Big[\sum_{\ell=1}^n\,\delta_{\ell (ab)}\,k_a\cdot {\partial\over\partial k_\ell}\,\Big({1\over s_{(i+1)(i+2)\cdots(j-1)j}}+{1\over s_{(j+1)(j+2)\cdots(i-1)i}}\Big)\Big]\,N_L(0)\,N_R(0).~~~~\label{D0}
\eea
It is worth giving some interpretation for the second equality of \eref{D0}. The momentum conservation indicates
\bea
{1\over s_{(i+1)(i+2)\cdots(j-1)j}}={1\over s_{(j+1)(j+2)\cdots(i-1)i}}\,.
\eea
For each $\ell\in\{i+1,\cdots,j\}$, we have
\bea
& &k_a\cdot {\partial\over\partial k_\ell}\,{1\over s_{(i+1)(i+2)\cdots(j-1)j}}=-{2k_a\cdot K_{(i+1)j}\over s^2_{(i+1)(i+2)\cdots(j-1)j}}\,,\nn
& &k_a\cdot {\partial\over\partial k_\ell}\,{1\over s_{(j+1)(j+2)\cdots(i-1)i}}=0\,,~~\label{case1}
\eea
while for $\ell\in\{j+1,\cdots,i\}$
\bea
& &k_a\cdot {\partial\over\partial k_\ell}\,{1\over s_{(j+1)(j+2)\cdots(i-1)i}}=-{2k_a\cdot K_{(j+1)i}\over s^2_{(i+1)(i+2)\cdots(j-1)j}}
={2k_a\cdot K_{(i+1)j}\over s^2_{(i+1)(i+2)\cdots(j-1)j}}\,,\nn
& &k_a\cdot {\partial\over\partial k_\ell}\,{1\over s_{(i+1)(i+2)\cdots(j-1)j}}=0\,.~~\label{case2}
\eea
Thus $\partial/\partial {k_\ell}$ for any $\ell\in\{i+1,\cdots,j\}\cup\{j+1,\cdots,i\}$ gives rise to the equivalent result, up to a sign. Here a subtle point is that the Mandelstam variable $s_{(i+1)(i+2)\cdots(j-1)j}$ contains $k_\ell^2=0$. Although $k_\ell^2$ vanish due to the on-shell condition, they contribute $2k_\ell^\mu$ when we take the derivative of $k_{\ell\mu}$. The symbol $\delta_{\ell (ab)}$ selects two of equivalent terms given by \eref{case1} and \eref{case2} among $\ell\in\{1,\cdots,n\}$, since there are only two non-zero $\delta_{\ell (ab)}$. The relative $-$ between \eref{case1} and \eref{case2} is absorbed into $\delta_{\ell (ab)}$. To see this, one can observe that for any $\pmb\sigma=\cdots, k,a,b,k+1,\cdots$ or $\pmb\sigma=\cdots, k,b,a,k+1,\cdots$, two effective $\delta_{\ell(ab)}$
are $\delta_{k(ab)}=1$ and $\delta_{(k+1)(ab)}=-1$, the relative $-$ between $\delta_{k(ab)}$ and $\delta_{(k+1)(ab)}$ cancels the relative $-$ between \eref{case1} and \eref{case2}.

In \eref{D0}, the signature ${\rm sign}(\pm)$ is now decomposed into ${\rm sign}'(\pm)$ and the sign carried by one of two effective $\delta_{\ell(ab)}$, i.e., $\delta_{k (ab)}$ for $\pmb\sigma=\cdots, k,a,b,k+1,\cdots$ or $\pmb\sigma=\cdots, k,b,a,k+1,\cdots$. The key observation is that ${\rm sign}'(\pm)$ determined in this way is the overall sign of the amplitude ${\cal A}_{\rm BAS}(1,\cdots,n||\pmb\sigma\setminus\pmb{2})$.
Consider the special case ${\cal A}_{\rm BAS}(1,\cdots,j,a,b,j+1,\cdots,n||1,\cdots,j,a,b,j+1,\cdots,n)$ with two identical orderings, the overall sign of this amplitude is $+$ due to our
convention. For this case, two nonzero $\delta_{\ell(ab)}$ are $\delta_{j(ab)}=1$ and $\delta_{(j+1)(ab)}=-1$. The associated ${\rm sign}'(\pm)$ is ${\rm sign}'(\pm)=1$, since the overall sign of the $n+2$-point amplitude is decomposed as
\bea
{\rm sign}({\cal A}_{n+2})={\rm sign}'(\pm)\delta_{ab}\delta_{j(ab)}\,,~~\label{deco}
\eea
and $\delta_{ab}=\delta_{j(ab)}=1$. This is just the overall sign for the amplitude ${\cal A}_{\rm BAS}(1,\cdots,n||1,\cdots,n)$. The general ordering $\pmb\sigma$ among $n+2$ legs can be generated from $(1,\cdots,j,a,b,j+1,\cdots,n)$ via permutations. For permutations which move $a$ or $b$, the relative $-$ created by flippings are absorbed into $\delta_{ab}$ and $\delta_{\ell(ab)}$. For permutations without moving $a$ and $b$, the corresponding $-$ are common for ${\cal A}_{\rm BAS}(1,\cdots,j,a,b,j+1,\cdots,n||\pmb\sigma)$ and ${\cal A}_{\rm BAS}(1,\cdots,n||\pmb\sigma\setminus\pmb{2})$. Consequently, the decomposition in \eref{deco} ensures that ${\rm sign}'(\pm)$ always serve as the overall sign of ${\cal A}_{\rm BAS}(1,\cdots,n||\pmb\sigma\setminus\pmb{2})$.

The above consideration holds for any internal line of ${\cal A}_{\rm BAS}(1,\cdots,n||\pmb\sigma\setminus\pmb{2})$ which is coupled to $1/s_{ab}$. Thus we can sum over corresponding $D^{(0)_{ab}}_{ij}$ to get the $\tau^{-1}$
contribution,
\bea
D&=&\,\sum_{i\in\{1,\cdots,n\}}\,\sum_{j\in\{1,\cdots,n\}\setminus i}\,D^{(0)_{ab}}_{ij}\nn
&=&-{\delta_{ab}\over 2\tau\, S_{ab}}\,\sum_{\ell=1}^n\,\delta_{\ell (ab)}\,k_a\cdot {\partial\over\partial k_\ell}\,{\cal A}_{\rm BAS}(1,\cdots,n||\pmb\sigma\setminus \pmb{2})\nn
&=&-{\delta_{ab}\over\tau\, s_{ab}}\,\sum_{\ell=1}^n\,{\delta_{\ell (ab)}\,\big(k_\ell\cdot k_{ab}\big)\over 2k_\ell\cdot k_{ab}}\,k_a\cdot {\partial\over\partial k_\ell}\,{\cal A}_{\rm BAS}(1,\cdots,n||\pmb\sigma\setminus \pmb{2})\,.~~~~\label{D-0}
\eea
Each internal line of ${\cal A}_{\rm BAS}(1,\cdots,n||\pmb\sigma\setminus\pmb{2})$ is summed twice, thus the factor $1/4$ in \eref{D0} is turned to $1/2$. The overall signature ${\rm sign}'(\pm)$ in \eref{D0} is absorbed by ${\cal A}_{\rm BAS}(1,\cdots,n||\pmb\sigma\setminus\pmb{2})$.

Combining \eref{B-1} and \eref{D-0} together gives
\bea
B_1+D=S^{(1)_{ab}}_{\rm diff}\,{\cal A}_{\rm BAS}(1,\cdots,n||\pmb\sigma\setminus \pmb{2})\,,
\eea
where
\bea
S^{(1)_{ab}}_{\rm diff}={\delta_{ab}\over\tau\, s_{ab}}\,\sum_{\ell=1}^n\,{\delta_{\ell (ab)}\,\over 2k_\ell\cdot k_{ab}}\,k_{\ell}\cdot L_{ab}\cdot {\partial\over\partial k_\ell}\,,~~\label{soft-diff}
\eea
with the tensor $L_{ab}$ defined as
\bea
L^{\mu\nu}_{ab}\equiv k^\mu_a\,k_{ab}^\nu-k_{ab}^\mu\,k_a^\nu=k_a^\mu\,k_b^\nu-k_b^\mu\,k_a^\nu\,.
\eea
The differential operator $S^{(1)_{ab}}_{\rm diff}$ is the only part which satisfies Leibnitz rule in the full double-soft factor.

The remaining part arises from Feynman diagrams those $a$ and $b$ couple to different vertices. For these diagrams, the $\tau^{-1}$ order is the leading order. Using the expansion formula \eref{2tr-general}, this part is found as
\bea
E={1\over\tau}\,\sum_{i=1}^{n-1}\Big({\delta_{ia}\,\delta_{bn}\over2s_{bn}}+(k_a\cdot Z_i)\,{\delta_{ai}\,\delta_{ib}\over s_{ai}\,s_{ib}}\Big)\,
{\cal A}_{\rm BAS}(1,\cdots,n||\pmb\sigma\setminus \pmb{2})\,,
\eea
where the combinatorial momenta $Z_i$ are defined as $Z_i\equiv\sum_{h=1}^{i-1}k_h$.
Thus we have
\bea
{\cal A}^{(1)_{ab}}_{\rm YMS}(\pmb{1}|\pmb{2}||\pmb\sigma)=S^{(1)_{ab}}_s\,{\cal A}_{\rm BAS}(\pmb{1}||\pmb\sigma\setminus\pmb{2})\,,
\eea
with
\bea
S^{(1)_{ab}}_s=S^{(1)_{ab}}_{\rm diff}+S^{(1)_{ab}}_{\rm fac}\,,
\eea
where $S^{(1)_{ab}}_{\rm fac}$ is obtained by combining $B_2$ and $E$, namely,
\bea
S^{(1)_{ab}}_{\rm fac}=-{\delta_{ab}\over \tau}\,\Big(\sum_{i=1}^n\,{\delta_{i(ab)}\,(k_i\cdot k_a)\over \big(2k_i\cdot k_{ab}\big)^2}\Big)
+{1\over\tau}\,\sum_{i=1}^{n-1}\Big({\delta_{ia}\,\delta_{bn}\over2s_{bn}}+(k_a\cdot Z_i)\,{\delta_{ai}\,\delta_{ib}\over s_{ai}\,s_{ib}}\Big)\,.
\eea
The factor $S^{(1)_{ab}}_{\rm fac}$ is not a differential operator.

\subsection{More length-$2$ traces}
\label{subsec-multi-length2}

In this subsection, we construct multi-trace YMS amplitudes which contain only length-2 traces and no gluon, by using the double-soft theorem established in the previous subsection.

We begin with ${\cal A}_{\rm YMS}(\pmb{1}|\pmb{2}|\pmb{3}||\pmb\sigma)$ where the three scalar traces are  $\pmb{1}=\{1,2\}$, $\pmb{2}=\{a,b\}$, $\pmb{3}=\{c,d\}$, respectively.
The double-soft theorem indicates
\bea
{\cal A}^{(1)_{cd}}_{\rm YMS}(\pmb{1}|\pmb{2}|\pmb{3}||\pmb\sigma)&=&S^{(1)_{cd}}_s\,{\cal A}_{\rm YMS}(\pmb{1}|\pmb{2}||\pmb\sigma\setminus\pmb{3})\nn
&=&S^{(1)_{cd}}_s\,\Big[(k_a\cdot k_1)\,{\cal A}_{\rm BAS}(1,a,b,2||\pmb\sigma\setminus\pmb{3})\Big]\nn
&=&(k_a\cdot k_1)\,\Big[S^{(1)_{cd}}_s\,{\cal A}_{\rm BAS}(1,a,b,2||\pmb\sigma\setminus\pmb{3})\Big]+\Big[S^{(1)_{cd}}_{\rm diff}\,(k_a\cdot k_1)\Big]\,{\cal A}_{\rm BAS}(1,a,b,2||\pmb\sigma\setminus\pmb{3})\nn
&=&(k_a\cdot k_1)\,{\cal A}^{(1)_{cd}}_{\rm YMS}(1,a,b,2|\pmb{3}||\pmb\sigma)+\Big[S^{(1)_{cd}}_{\rm diff}\,(k_a\cdot k_1)\Big]\,{\cal A}_{\rm BAS}(1,a,b,2||\pmb\sigma\setminus\pmb{3})\,.~~\label{3tr-1}
\eea
In the above, we have used the expansion formula \eref{2tr-4p} for the second equality. On the third line, the Lebniz rule for the differential operator $S^{(1)_{cd}}_{\rm diff}$ was applied. On the last line, we applied the double-soft theorem to obtain the first term, while the second term is further reduced as
\bea
& &\left[\,S^{(1)_{cd}}_{\rm diff}\,(k_a\cdot k_1)\right]\,{\cal A}_{\rm BAS}(1,a,b,2||\pmb\sigma\setminus\pmb{3})\label{3tr-diff}\\
&=&{\delta_{cd}\over \tau\,s_{cd}}\,\left[\Big({-\delta_{1(cd)}\over 2k_1\cdot k_{cd}}+{\delta_{a(cd)}\over 2k_a\cdot k_{cd}}\Big)\,(k_a\cdot L_{cd}\cdot k_1)\right]\,{\cal A}_{\rm BAS}(1,a,b,2||\pmb\sigma\setminus\pmb{3})\nn
&=&{1\over\tau\, s_{cd}}\,\left[\Big({\delta_{1(cd)}\over 2k_1\cdot k_{cd}}+{\delta_{(cd)a}\over 2k_a\cdot k_{cd}}\Big)\,\delta_{cd}\,(k_a\cdot k_d)(k_c\cdot k_1)\right]\,{\cal A}_{\rm BAS}(1,a,b,2||\pmb\sigma\setminus\pmb{3})\nn
& &+{1\over\tau\, s_{cd}}\,\Big[\Big({\delta_{1(cd)}\over 2k_1\cdot k_{cd}}+{\delta_{(cd)a}\over 2k_a\cdot k_{cd}}\Big)\,\delta_{dc}\,(k_a\cdot k_c)(k_d\cdot k_1)\Big]\,{\cal A}_{\rm BAS}(1,a,b,2||\pmb\sigma\setminus\pmb{3})\nn
&=&(k_a\cdot k_d)^{(0)_{cd}}\,(k_c\cdot k_1)^{(0)_{cd}}\,{\cal A}^{(0)_{cd}}_{\rm BAS}(1,c,d,a,b,2||\pmb\sigma)+(k_a\cdot k_c)^{(0)_{cd}}\,(k_d\cdot k_1)^{(0)_{cd}}\,{\cal A}^{(0)_{cd}}_{\rm BAS}(1,d,c,a,b,2||\pmb\sigma)\,.\nonumber~~
\eea
In the above calculation, the second equality is obtained by using the explicit formula of $S^{(1)_{cd}}_{\rm diff}$ in \eref{soft-diff}, as well as the anti-symmetry of $L^{\mu\nu}_{cd}$. The last line uses the double-soft theorem for pure BAS amplitudes in \eref{soft-2s1} and \eref{soft-fac-2s}. Substituting \eref{3tr-diff} into \eref{3tr-1} we get
\bea
{\cal A}^{(1)_{cd}}_{\rm YMS}(\pmb{1}|\pmb{2}|\pmb{3}||\pmb\sigma)&=&(k_a\cdot k_1)\,{\cal A}^{(1)_{cd}}_{\rm YMS}(1,a,b,2|\pmb{3}||\pmb\sigma)\nn
& &+(k_a\cdot k_d)^{(0)_{cd}}\,(k_c\cdot k_1)^{(0)_{cd}}\,{\cal A}^{(0)_{cd}}_{\rm BAS}(1,c,d,a,b,2||\pmb\sigma)\nn
& &+(k_a\cdot k_c)^{(0)_{cd}}\,(k_d\cdot k_1)^{(0)_{cd}}\,{\cal A}^{(0)_{cd}}_{\rm BAS}(1,d,c,a,b,2||\pmb\sigma)\,,~~\label{3tr-2}
\eea
which indicates the following expansion formula
\bea
{\cal A}_{\rm YMS}(\pmb{1}|\pmb{2}|\pmb{3}||\pmb\sigma)&=&(k_a\cdot k_1)\,{\cal A}_{\rm YMS}(1,a,b,2|\pmb{3}||\pmb\sigma)\nn
& &+(k_a\cdot k_d)\,(k_c\cdot k_1)\,{\cal A}_{\rm BAS}(1,c,d,a,b,2||\pmb\sigma)\nn
& &+(k_a\cdot k_c)\,(k_d\cdot k_1)\,{\cal A}_{\rm BAS}(1,d,c,a,b,2||\pmb\sigma)\,.~~\label{3tr-expan}
\eea

Inspired by the above discussion, we now insert more traces, each of which contains two scalars, by the double-soft behavior. The expansion formula \eqref{Eq:MultiPureScalar} for those amplitudes containing only length-$2$ traces is explicitly written as
\bea
&&{\cal A}_{\rm YMS}(\pmb{1}|\cdots|\pmb{m}||\pmb\sigma)~~\label{ktr-expan}\\
&=&\sum_{\substack{\pmb{t}_1,...,\pmb{t}_s\\ \in\pmb{\rm Tr}\setminus\{\pmb{1},\pmb{2}\}}}\,\underset{\substack{c_i,d_i\in\pmb{t}_i\\ \text{for~} i=1,...,s}}{\widetilde{\sum}} (k_a\cdot k_{d_1})(k_{c_1}\cdot k_{d_2})\cdots(k_{c_s}\cdot k_1)\,
{\cal A}_{\rm YMS}(1,c_s,d_s,\cdots,c_1,d_1,a,b,2|\pmb{u}_1|\cdots|\pmb{u}_{m-s-2}||\pmb\sigma)\,.\nonumber
\eea
Obviously, the examples \eref{2tr-4p} and \eref{3tr-expan} satisfy the general pattern in \eref{ktr-expan}. Now we use our recursive technic to prove that if the expansion \eref{ktr-expan} is satisfied by $k$-trace amplitudes,  it also holds for the $(k+1)$-trace ones. Let us encode elements in $\pmb{k+1}$ as $\pmb{k+1}=\{g,h\}$. The double-soft theorem requires
\bea
& &{\cal A}^{(1)_{gh}}_{\rm YMS}(\pmb{1}|\cdots|\pmb{k+1}||\pmb\sigma)\nn
&=&S^{(1)_{gh}}_s\,{\cal A}_{\rm YMS}\left(\pmb{1}|\cdots|\pmb{k}||\pmb\sigma\setminus\pmb{k+1}\right)\nn
&=&S^{(1)_{gh}}_s\,\Biggl[\,\sum_{\substack{\pmb{t}_1,...,\pmb{t}_s\\\in\pmb{\rm Tr}\setminus\{\pmb{1},\pmb{2},\pmb{k+1}\}}}\,\underset{\substack{c_i,d_i\in\pmb{t}_i\\ \text{for~} i=1,...,s}}{\widetilde{\sum}}\,C_{c,d}(\pmb{K})\,
{\cal A}_{\rm YMS}(1,c_s,d_s,\cdots,c_1,d_1,a,b,2|\pmb{u}_1|\cdots|\pmb{u}_{k-s-2}||\pmb\sigma\setminus\pmb{k+1})\Biggr]\nn
&=&\sum_{\substack{\pmb{t}_1,...,\pmb{t}_s\\\in\pmb{\rm Tr}\setminus\{\pmb{1},\pmb{2},\pmb{k+1}\}}}\,\underset{\substack{c_i,d_i\in\pmb{t}_i\\ \text{for~} i=1,...,s}}{\widetilde{\sum}}
\,\Biggl\{\,C_{c,d}(\pmb{K})\,\biggl[S^{(1)_{gh}}_s\,
{\cal A}_{\rm YMS}(1,c_s,d_s,\cdots,c_1,d_1,a,b,2|\pmb{u}_1|\cdots|\pmb{u}_{k-s-2}||\pmb\sigma\setminus\pmb{k+1})\biggr]\nn
& &~~~~~~~~~~~~~+\,\biggl[S^{(1)_{gh}}_{\rm diff}\,C_{c,d}(\pmb{K})\biggr]\,
{\cal A}_{\rm YMS}(1,c_s,d_s,\cdots,c_1,d_1,a,b,2|\pmb{u}_1|\cdots|\pmb{u}_{k-s-2}||\pmb\sigma\setminus\pmb{k+1})\,\Biggr\}.~~\label{k+1tr-1}
\eea
The first term on the last line of \eref{k+1tr-1} can be expressed by using the double-soft theorem
\bea
& &S^{(1)_{gh}}_s\,
{\cal A}_{\rm YMS}(1,c_s,d_s,\cdots,c_1,d_1,a,b,2|\pmb{u}_1|\cdots|\pmb{u}_{k-s-2}||\pmb\sigma\setminus\pmb{k+1})\nn
&=&{\cal A}^{(1)_{gh}}_{\rm YMS}(1,c_s,d_s,\cdots,c_1,d_1,a,b,2|\pmb{u}_1|\cdots|\pmb{u}_{k-s-2}||\pmb\sigma\setminus\pmb{k+1}).~~\label{part1}
\eea
When the definition of $S^{(1)_{gh}}_{\rm diff}$ in \eref{soft-diff} is applied, the coefficient in the second term of \eqref{k+1tr-1} reads
\bea
S^{(1)_{gh}}_{\rm diff}\,C_{c,d}(\pmb{K})&=&{\delta_{gh}\over\tau\,s_{gh}}\,\Biggl[\Big({\delta_{a(gh)}\over 2k_a\cdot k_{gh}}-{\delta_{c_1(gh)}\over 2k_{c_1}\cdot k_{gh}}\Big)\,\Big(k_a\cdot L_{gh}\cdot T_{\rho_1}\cdot T_{\rho_2}\cdots T_{\rho_s}\cdot k_1\Big)\nn
& &~~~~~~+\Big({\delta_{d_s(gh)}\over 2k_{d_s}\cdot k_{gh}}-{\delta_{1(gh)}\over 2k_{1}\cdot k_{gh}}\Big)\,\Big(k_a\cdot T_{\rho_1}\cdot T_{\rho_2}\cdots T_{\rho_s}\cdot L_{gh}\cdot k_1\Big)\nn
& &~~~~~~+\sum_{p=1}^{s-1}\,\Big({\delta_{d_p(gh)}\over 2k_{d_p}\cdot k_{gh}}-{\delta_{c_{p+1}(gh)}\over 2k_{c_{p+1}}\cdot k_{gh}}\Big)\,\Big(k_a\cdot T_{\rho_1}\cdots T_{\rho_p}\cdot L_{gh}\cdot T_{\rho_{p+1}}\cdots T_{\rho_s}\cdot k_1\Big)\Biggr]\nn
&=&{1\over\tau\,s_{gh}}\,\biggl[\delta_{hg}\,\Big({\delta_{(gh)a}\over 2k_a\cdot k_{gh}}+{\delta_{c_1(gh)}\over 2k_{c_1}\cdot k_{gh}}\Big)\,\Big(k_a\cdot T_{gh}\cdot T_{\rho_1}\cdot T_{\rho_2}\cdots T_{\rho_s}\cdot k_1\Big)\nn
& &~~~~~~+\delta_{hg}\,\Big({\delta_{(gh)d_s}\over 2k_{d_s}\cdot k_{gh}}+{\delta_{1(gh)}\over 2k_{1}\cdot k_{gh}}\Big)\,\Big(k_a\cdot T_{\rho_1}\cdot T_{\rho_2}\cdots T_{\rho_s}\cdot T_{gh}\cdot k_1\Big)\nn
& &~~~~~~+\sum_{p=1}^{s-1}\,\delta_{hg}\,\Big({\delta_{(gh)d_p}\over 2k_{d_p}\cdot k_{gh}}+{\delta_{c_{p+1}(gh)}\over 2k_{c_{p+1}}\cdot k_{gh}}\Big)\,\Big(k_a\cdot T_{\rho_1}\cdots T_{\rho_p}\cdot T_{gh}\cdot T_{\rho_{p+1}}\cdots T_{\rho_s}\cdot k_1\Big)\biggr]\nn
& &+(g\leftrightarrow h)\,,
\eea
where we have used the definitions $T^{\mu\nu}_{gh}\equiv k^\mu_gk^\nu_h$ and $L^{\mu\nu}_{gh}\equiv T^{\mu\nu}_{gh}-T^{\mu\nu}_{hg}$, as well as the antisymmetry of the operator $\delta_{gh}$. When the above expression is considered,  the second term of the third line in \eqref{k+1tr-1} turns into
\bea
& &\Big[S^{(1)_{gh}}_{\rm diff}\,C_{c,d}(\pmb{K})\Big]\,
{\cal A}_{\rm YMS}(1,c_s,d_s,...,c_1,d_1,a,b,2|\pmb{u}_1|\cdots|\pmb{u}_{k-s-2}||\pmb\sigma\setminus\pmb{k+1})\nn
&=&C^{(0)_{gh}}_{c,d}(\pmb{K})\,{\cal A}^{(0)_{gh}}_{\rm YMS}(1,\{c_s,d_s,...,c_1,d_1\}\W\shuffle\{h,g\},a,b,2|\pmb{u}_1|\cdots|\pmb{u}_{k-s-2}||\pmb\sigma)\,.~~\label{part2}
\eea
Substituting \eref{part1} and \eref{part2} into \eref{k+1tr-1}, we see that the expansion \eref{ktr-expan} holds for the $(k+1)$-trace case.

\subsection{General multi-trace YMS amplitude without external gluon}
\label{subsec-multi-nogluon}

In this subsection, we construct pure scalar multi-trace  amplitudes which involve more scalars in each trace.

First, one can  construct the expansion formula of amplitude ${\cal A}_{\rm YMS}(\pmb{1}|\cdots|\pmb{m}||\pmb\sigma)$ where the trace $\pmb{1}=\{1,\cdots,n\}$ includes more scalars and each of the other traces still contains two scalars, by single-soft behavior, as we have done in section. \ref{subsec-2trace-1more}. The explicit expansion formula is then presented by
\bea
& &{\cal A}_{\rm YMS}(\pmb{1}|\cdots|\pmb{m}||\pmb\sigma)~~\label{mtr-expan-1}\\
&=&\sum_{\substack{\pmb{K}(\pmb{\rm Tr}_s,c,d)\\\pmb{\rm Tr}_s\subset\pmb{\rm Tr}\setminus\{\pmb{1},\pmb{2}\}}}\,\underset{c_i,d_i\in\pmb{t}_i}{\widetilde{\sum}}\,{C_{c,d}(\pmb{K})}\,
{\cal A}_{\rm YMS}(1,\{2,\cdots,n-1\}\shuffle\{\pmb{K}(\pmb{\rm Tr}_s,c,d),a,b\},n|\pmb{u}_1|\cdots|\pmb{u}_{m-s-2}||\pmb\sigma)\,,\nonumber
\eea
where the coefficient $C_{c,d}(\pmb{K})$ is given by
\bea
C_{c,d}(\pmb{K})=k_a\cdot T_{\rho_1}\cdots T_{\rho_s}\cdot Y_{\rho_s}\,.~~\label{defin-C-3}
\eea

Second, we use the leading single-soft theorem for BAS scalar to construct ${\cal A}_{\rm YMS}(\pmb{1}|\cdots|\pmb{m}||\pmb\sigma)$ where the trace $\pmb{2}$ includes more than two elements. Let us consider $\pmb{2}=\{a,b,e\}$, and choose $e$ as the fiducial scalar.
The leading order soft theorem for $k_b\to\tau k_b$ requires
\bea
&&{\cal A}^{(0)_b}_{\rm YMS}(\pmb{1}|\cdots|\pmb{m}||\pmb\sigma)\nn
%
&=&\sum_{\substack{\pmb{K}(\pmb{\rm Tr}_s,c,d)\\\pmb{\rm Tr}_s\subset\pmb{\rm Tr}\setminus\{\pmb{1},\pmb{2}\}}}\underset{c_i,d_i\in\pmb{t}_i}{\widetilde{\sum}}C_{c,d}(\pmb{K})\bigg[S^{(0)_b}_s\,{\cal A}_{\rm YMS}\big(1,\{2,\cdots,n-1\}\shuffle\{\pmb{K}(\pmb{\rm Tr}_s,c,d),a,e\},n|\pmb{u}_1|\cdots|\pmb{u}_{m-s-2}||\pmb\sigma\setminus b\big)\bigg]\nn
&=&\sum_{\substack{\pmb{K}(\pmb{\rm Tr}_s,c,d)\\\pmb{\rm Tr}_s\subset\pmb{\rm Tr}\setminus\{\pmb{1},\pmb{2}\}}}\underset{c_i,d_i\in\pmb{t}_i}{\widetilde{\sum}}C_{c,d}(\pmb{K})\bigg[\Big({\delta_{ab}\over s_{ab}}+{\delta_{be}\over s_{be}}\Big){\cal A}_{\rm YMS}\big(1,\{2,\cdots,n-1\}\nn
&&~~~~~~~~~~~~~~~~~~~~~~~~~~~~~~~~~~~~~~~~~~~~~~~~~~~~~~~~~~~~\shuffle\{\pmb{K}(\pmb{\rm Tr}_s,c,d),a,e\},n|\pmb{u}_1|\cdots|\pmb{u}_{m-s-2}||\pmb\sigma\setminus b\big)\bigg]\nn
&=&\sum_{\substack{\pmb{K}(\pmb{\rm Tr}_s,c,d)\\\pmb{\rm Tr}_s\subset\pmb{\rm Tr}\setminus\{\pmb{1},\pmb{2}\}}}\underset{c_i,d_i\in\pmb{t}_i}{\widetilde{\sum}}C_{c,d}(\pmb{K}){\cal A}^{(0)_b}_{\rm YMS}\big(1,\{2,\cdots,n-1\}\shuffle\{\pmb{K}(\pmb{\rm Tr}_s,c,d),a,b,e\},n|\pmb{u}_1|\cdots|\pmb{u}_{m-s-2}||\pmb\sigma\big)\,,
\eea
where the \eref{tech} has been applied on the last line. The above expression implies the following term
\bea
& &\sum_{\substack{\pmb{K}(\pmb{\rm Tr}_s,c,d)\\\pmb{\rm Tr}_s\subset\pmb{\rm Tr}\setminus\{\pmb{1},\pmb{2}\}}}\underset{c_i,d_i\in\pmb{t}_i}{\widetilde{\sum}}C_{c,d}(\pmb{K}){\cal A}_{\rm YMS}\big(1,\{2,\cdots,n-1\}\shuffle\{\pmb{K}(\pmb{\rm Tr}_s,c,d),a,b,e\},n|\pmb{u}_1|\cdots|\pmb{u}_{m-s-2}||\pmb\sigma\big)~~\label{piece1}
\eea
in the expansion of ${\cal A}_{\rm YMS}(\pmb{1}|\cdots|\pmb{m}||\pmb\sigma)$, where $C_{c,d}(\pmb{K})=k_a\cdot T_{\rho_1}\cdots T_{\rho_s}\cdot Y_{\rho_s}$.
This corresponds to the contribution where elements in the trace $\pmb{2}$ have the relative order $a,b,e$.

Moreover, it is possible that the full expansion contains contributions from other relative order of the trace  $\pmb{2}$, at higher order of $\tau$ when $k_b\to \tau k_b$. To determine this part, one can consider the soft behavior of another particle $a$, i.e.,  $k_a\to\tau k_a$. In this soft limit, the amplitude becomes
\bea
&&{\cal A}^{(0)_a}_{\rm YMS}(\pmb{1}|\cdots|\pmb{m}||\pmb\sigma)\nn
%
&=&\sum_{\substack{\pmb{K}(\pmb{\rm Tr}_s,c,d)\\\pmb{\rm Tr}_s\subset\pmb{\rm Tr}\setminus\{\pmb{1},\pmb{2}\}}}\underset{c_i,d_i\in\pmb{t}_i}{\widetilde{\sum}}C_{c,d}(\pmb{K})\bigg[S^{(0)_a}_s\,{\cal A}_{\rm YMS}(1,\{2,\cdots,n-1\}\shuffle\{\pmb{K}(\pmb{\rm Tr}_s,c,d),b,e\},n|\pmb{u}_1|\cdots|\pmb{u}_{m-s-2}||\pmb\sigma\setminus a)\bigg]\nn
&=&\sum_{\substack{\pmb{K}(\pmb{\rm Tr}_s,c,d)\\\pmb{\rm Tr}_s\subset\pmb{\rm Tr}\setminus\{\pmb{1},\pmb{2}\}}}\underset{c_i,d_i\in\pmb{t}_i}{\widetilde{\sum}}C_{c,d}(\pmb{K})\bigg[-\Big({\delta_{ba}\over s_{ba}}+{\delta_{ae}\over s_{ae}}\Big){\cal A}_{\rm YMS}\Big(1,\{2,\cdots,n-1\}\nn
&&~~~~~~~~~~~~~~~~~~~~~~~~~~~~~~~~~~~~~~~~~~~~~~~~~~~~~~~~~~~~~~\shuffle\{\pmb{K}(\pmb{\rm Tr}_s,c,d),b,e\},n|\pmb{u}_1|\cdots|\pmb{u}_{m-s-2}||\pmb\sigma\setminus a\Big)\bigg]\nn
&=&-\sum_{\substack{\pmb{K}(\pmb{\rm Tr}_s,c,d)\\\pmb{\rm Tr}_s\subset\pmb{\rm Tr}\setminus\{\pmb{1},\pmb{2}\}}}\underset{c_i,d_i\in\pmb{t}_i}{\widetilde{\sum}}C_{c,d}(\pmb{K}){\cal A}^{(0)_a}_{\rm YMS}(1,\{2,\cdots,n-1\}\shuffle\{\pmb{K}(\pmb{\rm Tr}_s,c,d),b,a,e\},n|\pmb{u}_1|\cdots|\pmb{u}_{m-s-2}||\pmb\sigma)\,,
\eea
where the antisymmetry of $\delta_{ab}$ has been applied. The above implies the contribution corresponding to the relative order $b,a,e$:
\bea
& &-\sum_{\substack{\pmb{K}(\pmb{\rm Tr}_s,c,d)\\\pmb{\rm Tr}_s\subset\pmb{\rm Tr}\setminus\{\pmb{1},\pmb{2}\}}}\underset{c_i,d_i\in\pmb{t}_i}{\widetilde{\sum}}C_{c,d}(\pmb{K}){\cal A}_{\rm YMS}(1,\{2,\cdots,n-1\}\shuffle\{\pmb{K}(\pmb{\rm Tr}_s,c,d),b,a,e\},n|\pmb{u}_1|\cdots|\pmb{u}_{m-s-2}||\pmb\sigma).\nn\label{piece2}
\eea
The coefficient $C_{c,d}(\pmb{K})$ in the above is  $C_{c,d}(\pmb{K})=k_b\cdot T_{\rho_1}\cdots T_{\rho_s}\cdot Y_{\rho_1}$.

Combining the contributions corresponding to distinct relative orders, \eref{piece1} and \eref{piece2} together, we find the following expansion formula
\bea
& &~~~{\cal A}_{\rm YMS}(\pmb{1}|\cdots|\pmb{m}||\pmb\sigma)~~\label{multi-trace-scalar-expan}\\
&=&\sum_{\substack{\pmb{K}\big(\pmb{\rm Tr}_s,c,d)\\\pmb{\rm Tr}_s\subset\pmb{\rm Tr}\setminus\{\pmb{1},\pmb{2}\}}}\underset{c_i,d_i\in\pmb{t}_i}{\widetilde{\sum}}\underset{d_0\in\pmb{2} \setminus e}{\widetilde{\sum}}\,C_{c,d}(\pmb{K}){\cal A}_{\rm YMS}(1,\{2,\cdots,n-1\}\shuffle\{\pmb{K}(\pmb{\rm Tr}_s,c,d),K^{\pmb{2}}_{d_0,e}\},n|\pmb{u}_1|\cdots|\pmb{u}_{m-s-2}||\pmb\sigma\big)\,,\nonumber
\eea
in which, the coefficient is
\bea
C_{c,d}(\pmb{K})=k_{d_0}\cdot T_{\rho_1}\cdots T_{\rho_s}\cdot Y_{\rho_s}\,.~~\label{define-C}
\eea
The ordered set $K^{\pmb{2}}_{d_0,e}$ is defined as $K^{\pmb{2}}_{d_0,e}=\{a,b,e\}$ when $d_0=a$, while $K^{\pmb{2}}_{d_0,e}=\{b,a,e\}$ when $d_0=b$. In the latter case, the notation $\underset{d_0\in\pmb{2} \setminus e}{\widetilde{\sum}}$ carries a $-$ sign.

For a general $\pmb{2}$ which includes the fiducial scalar $e$ and involves more than three elements, one can define $K^{\pmb{2}}_{d_0,e}$ as $K^{\pmb{2}}_{d_0,e}=\{d_0,\pmb{\a}\shuffle\pmb{\b}^T,e\}$ if the trace $\pmb{2}$ can be written as  $\pmb{2}=\{d_0,\pmb{\a},e,\pmb{\b}\}$. The notation $\underset{d_0\in\pmb{2} \setminus e}{\widetilde{\sum}}$ carries the $(-)^{n_{\pmb{\b}}}$ sign and among all elements in $\pmb{2} \setminus e$.
 One can use the technic parallel to the three element case, to obtain this more general amplitude recursively. Supposing \eref{multi-trace-scalar-expan} holds for $\pmb{2}=\{\pmb{\gamma},e\}$, we insert the external scalar $a$ to generate $\pmb{2}'=\{a,\pmb{\gamma},e\}$. When we consider the $a$ as the soft particle, $k_a\to\tau k_a$ yields
\bea
&&~~~~{\cal A}^{(0)_a}_{\rm YMS}\big(\pmb{1}|\pmb{2}'|\cdots|\pmb{m}||\pmb\sigma\big)~~\label{piece1-asoft}\\
&=&-\sum_{\substack{\pmb{K}(\pmb{\rm Tr}_s,c,d)\\\pmb{\rm Tr}_s\subset\pmb{\rm Tr}\setminus\{\pmb{1},\pmb{2}\}}}\,\underset{c_i,d_i\in\pmb{t}_i}{\widetilde{\sum}}\,\underset{d_0\in\pmb{2} \setminus e}{\widetilde{\sum}}C_{c,d}(\pmb{K})\bigg[\left({\delta_{r_1a}\over s_{r_1a}}+{\delta_{ae}\over s_{ae}}\right)\nn
& &~~~~~~~~~~~~~~~~~~~~~~~~~~~~~~~~~~~~~{\cal A}_{\rm YMS}\big(1,\{2,\cdots,n-1\}\shuffle\left\{\pmb{K}(\pmb{\rm Tr}_s,c,d),K^{\pmb{2}}_{d_0,e}\right\},n|\pmb{u}_1|\cdots|\pmb{u}_{m-s-2}||\pmb\sigma\setminus a\big)\bigg]\nn
&=&-\sum_{\substack{\pmb{K}(\pmb{\rm Tr}_s,c,d)\\\pmb{\rm Tr}_s\subset\pmb{\rm Tr}\setminus\{\pmb{1},\pmb{2}\}}}\underset{c_i,d_i\in\pmb{t}_i}{\widetilde{\sum}}\underset{d_0\in\pmb{2} \setminus e}{\widetilde{\sum}}C_{c,d}(\pmb{K}){\cal A}^{(0)_a}_{\rm YMS}\big(1,\{2,\cdots,n-1\}\shuffle\left\{\pmb{K}(\pmb{\rm Tr}_s,c,d),K^{\pmb{2}}_{d_0,e,a}\right\},n|\pmb{u}_1|\cdots|\pmb{u}_{m-s-2}||\pmb\sigma\big).\nonumber
\eea
For the second equality, we note that in the above expression, $\pmb{\gamma}=\left\{\pmb{\beta},d_0,\pmb{\alpha}\right\}$ and $r_1$ stands for the first element of $\pmb{\gamma}$, in other words, the first element of $\pmb{\beta}$.  The ordered set $K^{\pmb{2}}_{d_0,e,a}$ on the last line  is
\bea
K^{\pmb{2}}_{d_0,e,a}=\left\{d_0,\pmb{\a}\shuffle\{a,\pmb{\b}\}^T,e\right\}\,.
\eea
The soft behavior  in \eref{piece1-asoft} finally indicates the pieces with the leftmost element $d_0\neq a$ of $\pmb{2}$ in the expansion formula of ${\cal A}_{\rm YMS}(\pmb{1}|\cdots|\pmb{m}||\pmb\sigma)$
\bea
& &-\sum_{\substack{\pmb{K}(\pmb{\rm Tr}_s,c,d)\\\pmb{\rm Tr}_s\subset\pmb{\rm Tr}\setminus\{\pmb{1},\pmb{2}\}}}\underset{c_i,d_i\in\pmb{t}_i}{\widetilde{\sum}}\underset{d_0\in\pmb{2} \setminus e}{\widetilde{\sum}}\,C_{c,d}(\pmb{K}){\cal A}_{\rm YMS}\left(1,\{2,\cdots,n-1\}\shuffle\big\{\pmb{K}(\pmb{\rm Tr}_s,c,d),K^{\pmb{2}}_{d_0,e,a}\big\},n|\pmb{u}_1|\cdots|\pmb{u}_{m-s-2}||\pmb\sigma\right)~~\label{p1}\nn
\eea
To induce the terms where the scalar $a$ plays as $d_0$,  we should consider the soft behavior of any other $k_{r_i}$ where the trace $\pmb{2}$ has the form $K^{\pmb{2}}_{a,e}$. This soft behavior leads to
\bea
& &\sum_{\substack{\pmb{K}(\pmb{\rm Tr}_s,c,d)\\\pmb{\rm Tr}_s\subset\pmb{\rm Tr}\setminus\{\pmb{1},\pmb{2}\}}}\underset{c_i,d_i\in\pmb{t}_i}{\widetilde{\sum}}C_{c,d}(\pmb{K}){\cal A}_{\rm YMS}\left(1,\{2,\cdots,n-1\}\shuffle\left\{\pmb{K}(\pmb{\rm Tr}_s,c,d),K^{\pmb{2}}_{a,e}\right\},n|\pmb{u}_1|\cdots|\pmb{u}_{m-s-2}||\pmb\sigma\right)\,.~~\label{p2}
\eea
Combining \eref{p1} and \eref{p2} together, we find that the expansion of ${\cal A}_{\rm YMS}(\pmb{1}|\pmb{2}'|\cdots|\pmb{m}||\pmb\sigma)$ also satisfies \eref{multi-trace-scalar-expan}.

Using the extremely similar technic, one can also extend other traces $\pmb{t}_i$ in $\pmb{K}(\pmb{\rm Tr}_s,c,d)$ to general ordered set with an arbitrary
length, by inserting external scalars. Hence we have proven the expansion formula \eqref{Eq:MultiPureScalar} via soft behaviors.

%
%
%
%

\section{Multi-trace YMS amplitudes with external gluons}
\label{sec-YMS-sg}

In this section, we insert external gluons to the multi-trace amplitudes with only scalars via single-soft behavior of gluon, and obtain the expansion formula for YMS amplitudes with arbitrary numbers of gluons and scalar traces.

%
%
%
%

\subsection{One external gluon}
\label{subsec-sg-1g}

We first derive the recursive expansion of the YMS amplitude ${\cal A}_{\rm YMS}(\pmb{1}|\pmb{2}|\cdots|\pmb{m};p||\pmb\sigma)$ which involves only one gluon $p$. When considering $p$ as the soft gluon with momentum $k_p\to \tau k_p$ ($\tau\to 0$), according to the soft gluon behavior (\ref{soft-impose}), (\ref{soft-fac-g-0-2}) and the expansion formula for amplitudes with only scalar traces, we have
\bea
{\cal A}^{(1)_p}_{\rm YMS}(\pmb{1}|\pmb{2}|\cdots|\pmb{m};p||\pmb\sigma)
%
&=&\sum_{\substack{\pmb{K}(\pmb{\rm Tr}_s,c,d)\\\pmb{\rm Tr}_s\subset\pmb{\rm Tr}\setminus\{\pmb{1},\pmb{2}\}}}\,\underset{c_i,d_i\in\pmb{t}_i}{\widetilde{\sum}}\,\underset{d_0\in\pmb{2} \setminus e}{\widetilde{\sum}}\,C_{c,d}(\pmb{K})\,\left[S^{(1)_p}_g\,{\rm A}\left(\pmb{\rm Tr}_s,e\right)\right]\nn
& &~~~~~~~~~~+\sum_{\substack{\pmb{K}(\pmb{\rm Tr}_s,c,d)\\\pmb{\rm Tr}_s\subset\pmb{\rm Tr}\setminus\{\pmb{1},\pmb{2}\}}}\underset{c_i,d_i\in\pmb{t}_i}{\widetilde{\sum}}\underset{d_0\in\pmb{2} \setminus e}{\widetilde{\sum}}\left[S^{(1)_p}_g\,\,C_{c,d}(\pmb{K})\right]{\rm A}\left(\pmb{\rm Tr}_s,e\right)\nn
&=&\sum_{\substack{\pmb{K}(\pmb{\rm Tr}_s,c,d)\\\pmb{\rm Tr}_s\subset\pmb{\rm Tr}\setminus\{\pmb{1},\pmb{2}\}}}\underset{c_i,d_i\in\pmb{t}_i}{\widetilde{\sum}}\underset{d_0\in\pmb{2} \setminus e}{\widetilde{\sum}}\,C_{c,d}(\pmb{K})\,{\rm A}^{(1)_p}(\pmb{\rm Tr}_s,e;p)\nn
& &~~~~~~~~~~+\sum_{\substack{\pmb{K}(\pmb{\rm Tr}_s,c,d)\\\pmb{\rm Tr}_s\subset\pmb{\rm Tr}\setminus\{\pmb{1},\pmb{2}\}}}\underset{c_i,d_i\in\pmb{t}_i}{\widetilde{\sum}}\underset{d_0\in\pmb{2} \setminus e}{\widetilde{\sum}}\,\left[S^{(1)_p}_g\,\,C_{c,d}(\pmb{K})\right]{\rm A}(\pmb{\rm Tr}_s,e)\,,~~~~\label{sub-1g-1}
\eea
where ${\rm A}(\pmb{\rm Tr}_s,e)$ is the following combination of amplitudes
\bea
{\rm A}(\pmb{\rm Tr}_s,e)={\cal A}_{\rm YMS}(1,\{2,\cdots,n-1\}\shuffle\{\pmb{K}(\pmb{\rm Tr}_s,c,d),K^{\pmb{2}}_{d_0,e}\},n|\pmb{u}_1|\cdots|\pmb{u}_{m-s-2}||\pmb\sigma)\,,
\eea
and ${\rm A}(\pmb{\rm Tr}_s,e;p)$ is the combination of amplitudes with the gluon $p$ added
\bea
{\rm A}(\pmb{\rm Tr}_s,e;p)={\cal A}_{\rm YMS}(1,\{2,\cdots,n-1\}\shuffle\{\pmb{K}(\pmb{\rm Tr}_s,c,d),K^{\pmb{2}}_{d_0,e}\},n|\pmb{u}_1|\cdots|\pmb{u}_{m-s-2};p||\pmb\sigma)\,,
\eea
Here we have used the expansion in \eref{multi-trace-scalar-expan} and Leibnitz's rule to get the second equality, and the soft theorem to get the third.

To continue, we need to treat the contribution from the last line in \eref{sub-1g-1}. Using the explicit formula of $C_{c,d}(\pmb{K})$ in \eref{define-C}, together with the properties \eref{iden-1} and \eref{iden-2}, we find
\bea
S^{(1)_p}_g\,C_{c,d}(\pmb{K})&=&\left({\delta_{c_1 p}\over s_{c_1p}}+{\delta_{pd_0}\over s_{pd_0}}\right)\,k_{d_0}\cdot f_p\cdot T_{\rho_1}\cdots T_{\rho_s}\cdot Y_{\rho_s}\nn
& &+\sum_{\ell=1}^{s-1}\,\left({\delta_{c_{\ell+1} p}\over s_{c_{\ell+1}p}}+{\delta_{pd_\ell}\over s_{pd_\ell}}\right)\,k_{d_0}\cdot T_{\rho_1}\cdots T_{\rho_\ell}\cdot f_p\cdot T_{\rho_{\ell+1}}\cdots T_{\rho_s}\cdot Y_{\rho_s}\nn
& &+\sum_{i=1}^j\,\left({\delta_{ip}\over s_{ip}}+{\delta_{pd_s}\over s_{pd_s}}\right)\,k_{d_0}\cdot T_{\rho_1}\cdots T_{\rho_s}\cdot f_p\cdot k_i\,,~~~~\label{s-c}
\eea
where $Y_{\rho_s}$ is assumed to be $Y_{\rho_s}=\sum_{i=1}^j k_i$,  $d_l$ and $c_{l+1}$ are last and first elements of ordered sets $\{c_\ell,\mathsf{KK}[\pmb{\ell},c_\ell,d_\ell],d_\ell\}$ and $\{c_{\ell+1},\mathsf{KK}[\pmb{\ell+1},c_{\ell+1},d_{\ell+1}],d_{\ell+1}\}$, respectively. The last line in the above expression
is further reorganized as
\bea
& &\sum_{i=1}^j\,\left({\delta_{ip}\over s_{ip}}+{\delta_{pd_s}\over s_{pd_s}}\right)\,k_{d_0}\cdot T_{\rho_1}\cdots T_{\rho_s}\cdot f_p\cdot k_i\nn
&=&\sum_{i=1}^j\,\left[\sum_{\ell=i}^{j-1}\,\left({\delta_{\ell p}\over s_{\ell p}}+{\delta_{p(\ell+1)}\over s_{p(\ell+1)}}\right)+{\delta_{jp}\over s_{jp}}+{\delta_{pd_s}\over s_{pd_s}}\right]\,k_{d_0}\cdot T_{\rho_1}\cdots T_{\rho_s}\cdot f_p\cdot k_i\nn
&=&\sum_{\ell=1}^j\,\left({\delta_{\ell p}\over s_{\ell p}}+{\delta_{p(\ell+1)}\over s_{p(\ell+1)}}\right)\,\left[k_{d_0}\cdot T_{\rho_1}\cdots T_{\rho_s}\cdot f_p\cdot \Big(\sum_{i=1}^\ell\,k_i\Big)\right]\,.~~\label{reform}
\eea
On the last line, the $j+1$ is understood as $j+1=d_s$. We have exchanged the summation over $\ell$ and the summation over $i$, by collecting terms for a given $\ell$ together. This manipulation in fact produces $\Big(\sum_{i=1}^\ell\,k_i\Big)$ for a given $\ell$. Substituting \eref{reform} into \eref{s-c}, and reform the first and the second line in
\eref{s-c} via the technic in \eref{tech}, we find
\bea
& &\sum_{\substack{\pmb{K}(\pmb{\rm Tr}_s,c,d)\\\pmb{\rm Tr}_s\subset\pmb{\rm Tr}\setminus\{\pmb{1},\pmb{2}\}}}\,\underset{c_i,d_i\in\pmb{t}_i}{\widetilde{\sum}}\,\underset{d_0\in\pmb{2} \setminus e}{\widetilde{\sum}}\,\left[S^{(1)_p}_g\,\,C_{c,d}\left(\pmb{K}\right)\right]\,{\rm A}\left(\pmb{\rm Tr}_s,e\right)\nn
&=&\tau\sum_{\substack{\pmb{K}(\pmb{\rm Tr}_s,c,d)\\\pmb{\rm Tr}_s\subset\pmb{\rm Tr}\setminus\{\pmb{1},\pmb{2}\}}}\,\underset{c_i,d_i\in\pmb{t}_i}{\widetilde{\sum}}\,\underset{d_0\in\pmb{2} \setminus e}{\widetilde{\sum}}\,\,C_{c,d}\left(\pmb{K}\W\shuffle p\right)\,{\rm A}^{(0)_p}\left(\pmb{\rm Tr}_s\W\shuffle p,e\right)\,,~~\label{sc-1}
\eea
where
\bea
{\rm A}\left(\pmb{\rm Tr}_s\W\shuffle p,e\right)={\cal A}_{\rm YMS}\left(1,\{2,\cdots,n-1\}\shuffle\left\{\pmb{K}\left(\pmb{\rm Tr}_s,c,d\right)\W\shuffle p,K^{\pmb{2}}_{d_0,e}\right\},n|\pmb{u}_1|\cdots|\pmb{u}_{m-s-2}||\pmb\sigma\right)\,,~~~\label{defin-A-1}
\eea
and coefficients $C_{c,d}(\pmb{K}\W\shuffle p)$ are given by
\bea
C_{c,d}(\pmb{K}\W\shuffle p)=k_{d_0}\cdot T_{\rho_1}\cdot T_{\rho_2}\cdots T_{\rho_{s+1}}\cdot Y_{\rho_{s+1}}\,,~~~\label{defin-C-1}
\eea
where we denoted the ordered set arises from $\pmb{K}\left(\pmb{\rm Tr}_s,c,d\right)\W\shuffle p$ as $\left\{\rho_1,\cdots,\rho_{s+1}\right\}$,
and the tensor $T_{\rho_r}^{\mu\nu}$ should be understood as
\bea
T_{\rho_r}^{\mu\nu}=\begin{cases}\displaystyle ~f_p^{\mu\nu}~~~~ & \rho_r=p\,,\\
\displaystyle ~k_{c_\ell}^\mu k_{d_\ell}^\nu~~~~~ & \rho_r=\{c_\ell,\mathsf{KK}[\pmb{\ell},c_\ell,d_\ell],d_\ell\}.\end{cases}
\eea

Substituting \eref{sc-1} into \eref{sub-1g-1}, we get
\bea
{\cal A}^{(1)}_{\rm YMS}(\pmb{1}|\pmb{2}|\cdots|\pmb{m};p||\pmb\sigma)
&=&\sum_{\substack{\pmb{K}(\pmb{\rm Tr}_s,c,d)\\\pmb{\rm Tr}_s\subset\pmb{\rm Tr}\setminus\{\pmb{1},\pmb{2}\}}}\,\underset{c_i,d_i\in\pmb{t}_i}{\widetilde{\sum}}\,\underset{d_0\in\pmb{2} \setminus e}{\widetilde{\sum}}\,C_{c,d}(\pmb{K})\,{\rm A}^{(0)_p}\left(\pmb{\rm Tr}_s,e;p\right)\nn
& &+\tau\,\sum_{\substack{\pmb{K}(\pmb{\rm Tr}_s,c,d)\\\pmb{\rm Tr}_s\subset\pmb{\rm Tr}\setminus\{\pmb{1},\pmb{2}\}}}\,\underset{c_i,d_i\in\pmb{t}_i}{\widetilde{\sum}}\,\underset{d_0\in\pmb{2} \setminus e}{\widetilde{\sum}}\,\,C_{c,d}(\pmb{K}\W\shuffle p)\,{\rm A}^{(0)_p}\left(\pmb{\rm Tr}_s\W\shuffle p,e\right)\,.~~~~\label{sub-1g-2}
\eea
Since $C_{c,d}(\pmb{K})\to C_{c,d}(\pmb{K})$ and $C_{c,d}(\pmb{K}\W\shuffle p)\to\tau C_{c,d}(\pmb{K}\W\shuffle p)$ when $k_p\to\tau k_p$, from \eref{sub-1g-2}
one can conclude the YMS amplitude ${\cal A}_{\rm YS}(\pmb{1}|\pmb{2}|\cdots|\pmb{m};g||\sigma)$ satisfies the expansion relation
\bea
{\cal A}_{\rm YMS}(\pmb{1}|\pmb{2}|\cdots|\pmb{m};p||\pmb\sigma)
&=&\sum_{\substack{\pmb{K}(\pmb{\rm Tr}_s,c,d)\\\pmb{\rm Tr}_s\subset\pmb{\rm Tr}\setminus\{\pmb{1},\pmb{2}\}}}\,\underset{c_i,d_i\in\pmb{t}_i}{\widetilde{\sum}}\,\underset{d_0\in\pmb{2} \setminus e}{\widetilde{\sum}}\,C_{c,d}(\pmb{K})\,{\rm A}(\pmb{\rm Tr}_s,e;p)\nn
& &+\sum_{\substack{\pmb{K}(\pmb{\rm Tr}_s,c,d)\\\pmb{\rm Tr}_s\subset\pmb{\rm Tr}\setminus\{\pmb{1},\pmb{2}\}}}\,\underset{c_i,d_i\in\pmb{t}_i}{\widetilde{\sum}}\,\underset{d_0\in\pmb{2} \setminus e}{\widetilde{\sum}}\,\,C_{c,d}(\pmb{K}\W\shuffle p)\,{\rm A}(\pmb{\rm Tr}_s\W\shuffle p,e)\,.~~~~\label{expan-1g}
\eea
%

\subsection{General case}
\label{subsec-sg-general}

The expansion of general multi-trace YMS amplitude with $h$ external gluons is given as
\bea
{\cal A}_{\rm YMS}\left(\pmb{1}|\pmb{2}|\cdots|\pmb{m};\{g_k\}||\pmb\sigma\right)&=&\underset{\pmb{\rm g}\subset\{g_k\}}{\sum}\,\sum_{\substack{\pmb{K}(\pmb{\rm Tr}_s,c,d)\\\pmb{\rm Tr}_s\subset\pmb{\rm Tr}\setminus\{\pmb{1},\pmb{2}\}}}\,\underset{c_i,d_i\in\pmb{t}_i}{\widetilde{\sum}}\,\underset{d_0\in\pmb{2} \setminus e}{\widetilde{\sum}}\,C_{c,d}(\pmb{K}\W\shuffle\pmb{\rm g})\,{\rm A}\left(\pmb{\rm Tr}_s\W\shuffle\pmb{\rm g},e;\{g_k\}\setminus\pmb{\rm g}\right)\,,~~~~\label{expan-hg}
\eea
where $k\in\{1,\cdots,h\}$, and $\pmb{\rm g}$ is an ordered set whose elements belong to $\{g_k\}$. The explicit formulas of $C_{c,d}(\pmb{K}\W\shuffle\pmb{\rm g})$ and ${\rm A}(\pmb{\rm Tr}_s\W\shuffle\pmb{\rm g},e;\{g_k\}\setminus\pmb{\rm g})$ are respectively presented as
\bea
C_{c,d}(\pmb{K}\W\shuffle\pmb{\rm g})=k_{d_0}\cdot T_{\rho_1}\cdot T_{\rho_2}\cdots T_{\rho_{s+h}}\cdot Y_{\rho_{s+h}}\,,~~~\label{defin-C-2}
\eea
and
\bea
{\rm A}(\pmb{\rm Tr}_s\W\shuffle \pmb{\rm g},e;\{g_k\}\setminus\pmb{\rm g})={\cal A}_{\rm YMS}(1,\{2,\cdots,n-1\}\shuffle\{\pmb{K}[\pmb{\rm Tr}_s,c,d]\W\shuffle \pmb{\rm g},K^{\pmb{2}}_{d_0,e}\},n|\pmb{u}_1|\cdots|\pmb{u}_{m-s-2}||\pmb\sigma)\,,~~~\label{defin-A-1}
\eea
where the ordered set $\{\rho_1,\cdots,\rho_{s+h}\}$ arises from $\pmb{K}\left(\pmb{\rm Tr}_s,a,b\right)\W\shuffle \pmb{\rm g}$.

The general expansion \eref{expan-hg} can be derived from \eref{expan-1g} recursively. The expansion \eref{expan-1g} satisfies the general formula \eref{expan-hg} manifestly, with $\pmb{\rm g}=p$ and $\pmb{\rm g}=\emptyset$. Thus, we only need to show that if \eref{expan-hg} is correct for the $h$-gluon case ${\cal A}_{\rm YMS}(\pmb{1}|\pmb{2}|\cdots|\pmb{m};\{g_k\}||\pmb\sigma\setminus g_0)$, then it is also correct for the $(h+1)$-gluon case ${\cal A}_{\rm YMS}(\pmb{1}|\pmb{2}|\cdots|\pmb{m};\{g_k\}\cup g_0||\pmb\sigma)$. The process is paralleled to that
from \eref{sub-1g-1} to \eref{expan-1g}.

Let us take $g_0$ as the soft gluon, i.e., $k_{g_0}\to\tau k_{g_0}$, and consider the soft behavior of ${\cal A}_{\rm YMS}(\pmb{1}|\pmb{2}|\cdots|\pmb{m};\{g_k\}\cup g_0||\pmb\sigma)$
at the subleading order. The soft theorem requires
\bea
& &{\cal A}_{\rm YMS}^{(1)_{g_0}}(\pmb{1}|\pmb{2}|\cdots|\pmb{m};\{g_k\}\cup g_0||\pmb\sigma)\nn
&=&\underset{\pmb{\rm g}\subset\{g_k\}}{\sum}\,\sum_{\substack{\pmb{K}(\pmb{\rm Tr}_s,c,d)\\\pmb{\rm Tr}_s\subset\pmb{\rm Tr}\setminus\{\pmb{1},\pmb{2}\}}}\,\underset{c_i,d_i\in\pmb{t}_i}{\widetilde{\sum}}\,\underset{d_0\in\pmb{2} \setminus e}{\widetilde{\sum}}\,C_{c,d}(\pmb{K}\W\shuffle\pmb{\rm g})\,\left[S^{(1)_{g_0}}_g\,{\rm A}(\pmb{\rm Tr}_s\W\shuffle\pmb{\rm g},e;\{g_k\}\setminus\pmb{\rm g})\right]\nn
& &~~~~+\underset{\pmb{\rm g}\subset\{g_k\}}{\sum}\,\sum_{\substack{\pmb{K}(\pmb{\rm Tr}_s,c,d)\\\pmb{\rm Tr}_s\subset\pmb{\rm Tr}\setminus\{\pmb{1},\pmb{2}\}}}\,\underset{c_i,d_i\in\pmb{t}_i}{\widetilde{\sum}}\,\underset{d_0\in\pmb{2} \setminus e}{\widetilde{\sum}}\,\left[S^{(1)_{g_0}}_g\,C_{c,d}(\pmb{K}\W\shuffle\pmb{\rm g})\right]\,{\rm A}\left(\pmb{\rm Tr}_s\W\shuffle\pmb{\rm g},e;\{g_k\}\setminus\pmb{\rm g}\right)\nn
&=&\underset{\pmb{\rm g}\subset\{g_k\}}{\sum}\,\sum_{\substack{\pmb{K}(\pmb{\rm Tr}_s,c,d)\\\pmb{\rm Tr}_s\subset\pmb{\rm Tr}\setminus\{\pmb{1},\pmb{2}\}}}\,\underset{c_i,d_i\in\pmb{t}_i}{\widetilde{\sum}}\,\underset{d_0\in\pmb{2} \setminus e}{\widetilde{\sum}}\,C_{c,d}(\pmb{K}\W\shuffle\pmb{\rm g})\,{\rm A}^{(1)_{g_0}}\left(\pmb{\rm Tr}_s\W\shuffle\pmb{\rm g},e;\{g_k\}\cup g_0\setminus\pmb{\rm g}\right)\nn
& &~~~~+\underset{\pmb{\rm g}\subset\{g_k\}}{\sum}\,\sum_{\substack{\pmb{K}(\pmb{\rm Tr}_s,c,d)\\\pmb{\rm Tr}_s\subset\pmb{\rm Tr}\setminus\{\pmb{1},\pmb{2}\}}}\,\underset{c_i,d_i\in\pmb{t}_i}{\widetilde{\sum}}\,\underset{d_0\in\pmb{2} \setminus e}{\widetilde{\sum}}\,\left[S^{(1)_{g_0}}_g\,C_{c,d}\left(\pmb{K}\W\shuffle\pmb{\rm g})\right]\,{\rm A}(\pmb{\rm Tr}_s\W\shuffle\pmb{\rm g},e;\{g_k\}\setminus\pmb{\rm g}\right)\,.~~~~\label{sub-hg-1}
\eea
Similar as in \eref{s-c} and \eref{reform}, it is straightforward to get
\bea
S^{(1)_{g_0}}_g\,C_{c,d}(\pmb{K})&=&\Big({\delta_{c_1 g_0}\over s_{c_1g_0}}+{\delta_{g_0d_0}\over s_{g_0d_0}}\Big)\,k_{d_0}\cdot f_{g_0}\cdot T_{\rho_1}\cdots T_{\rho_{s+h}}\cdot Y_{\rho_s}\nn
& &+\sum_{\ell=1}^{s-1}\,\Big({\delta_{c_{\ell+1} g_0}\over s_{c_{\ell+1}g_0}}+{\delta_{g_0d_\ell}\over s_{g_0d_\ell}}\Big)\,k_{d_0}\cdot T_{\rho_1}\cdots T_{\rho_\ell}\cdot f_{g_0}\cdot T_{\rho_{\ell+1}}\cdots T_{\rho_{s+h}}\cdot Y_{\rho_s}\nn
& &+\sum_{\ell=1}^j\,\Big({\delta_{\ell g_0}\over s_{\ell g_0}}+{\delta_{g_0(\ell+1)}\over s_{g_0(\ell+1)}}\Big)\,\Big[k_{d_0}\cdot T_{\rho_1}\cdots T_{\rho_{s+h}}\cdot f_{g_0}\cdot \Big(\sum_{i=1}^\ell\,k_i\Big)\Big]\,,
\eea
with $Y_{\rho_{s+h}}=\sum_{i=1}^j k_i$. Subsequently
\bea
\left[S^{(1)_{g_0}}_g\,C_{c,d}(\pmb{K}\W\shuffle\pmb{\rm g})\right]\,{\rm A}\left(\pmb{\rm Tr}_s\W\shuffle\pmb{\rm g},e;\{g_k\}\setminus\pmb{\rm g}\right)=\tau\,C_{c,d}(\pmb{K}\W\shuffle\pmb{\rm g}\W\shuffle g_0)\,{\rm A}^{(0)_{g_0}}\left(\pmb{\rm Tr}_s\W\shuffle\pmb{\rm g}\W\shuffle g_0,e;\{g_k\}\setminus\pmb{\rm g}\right)\,.~~~\label{sc-2}
\eea
Substituting \eref{sc-2} into \eref{sub-hg-1}, we obtain
\bea
& &{\cal A}^{(1)_{g_0}}_{\rm YMS}(\pmb{1}|\pmb{2}|\cdots|\pmb{m};\{g_k\}\cup g_0||\pmb\sigma)\nn
&=&\underset{\pmb{\rm g}\subset\{g_k\}}{\sum}\,\sum_{\substack{\pmb{K}(\pmb{\rm Tr}_s,c,d)\\\pmb{\rm Tr}_s\subset\pmb{\rm Tr}\setminus\{\pmb{1},\pmb{2}\}}}\,\underset{c_i,d_i\in\pmb{t}_i}{\widetilde{\sum}}\,\underset{d_0\in\pmb{2} \setminus e}{\widetilde{\sum}}\,C_{c,d}(\pmb{K}\W\shuffle\pmb{\rm g})\,{\rm A}^{(1)_{g_0}}\left(\pmb{\rm Tr}_s\W\shuffle\pmb{\rm g},e;\{g_k\}\cup g_0\setminus\pmb{\rm g}\right)\nn
& &~~~~~+\tau\,\underset{\pmb{\rm g}\subset\{g_k\}}{\sum}\,\sum_{\substack{\pmb{K}(\pmb{\rm Tr}_s,c,d)\\\pmb{\rm Tr}_s\subset\pmb{\rm Tr}\setminus\{\pmb{1},\pmb{2}\}}}\,\underset{c_i,d_i\in\pmb{t}_i}{\widetilde{\sum}}\,\underset{d_0\in\pmb{2} \setminus e}{\widetilde{\sum}}\,C_{c,d}(\pmb{K}\W\shuffle\pmb{\rm g}\W\shuffle g_0)\,{\rm A}^{(0)_{g_0}}\left(\pmb{\rm Tr}_s\W\shuffle\pmb{\rm g}\W\shuffle g_0,e;\{g_k\}\setminus\pmb{\rm g}\right)\,,
\eea
which indicates the following general expansion formula
\bea
& &{\cal A}_{\rm YMS}(\pmb{1}|\pmb{2}|\cdots|\pmb{m};\{g_k\}\cup g_0||\pmb\sigma)\nn
&=&\underset{\pmb{\rm g}\subset\{g_k\}}{\sum}\,\sum_{\substack{\pmb{K}(\pmb{\rm Tr}_s,c,d)\\\pmb{\rm Tr}_s\subset\pmb{\rm Tr}\setminus\{\pmb{1},\pmb{2}\}}}\,\underset{c_i,d_i\in\pmb{t}_i}{\widetilde{\sum}}\,\underset{d_0\in\pmb{2} \setminus e}{\widetilde{\sum}}\,C_{c,d}(\pmb{K}\W\shuffle\pmb{\rm g})\,{\rm A}\left(\pmb{\rm Tr}_s\W\shuffle\pmb{\rm g},e;\{g_k\}\cup g_0\setminus\pmb{\rm g}\right)\nn
& &~~~~~+\underset{\pmb{\rm g}\subset\{g_k\}}{\sum}\,\sum_{\substack{\pmb{K}(\pmb{\rm Tr}_s,c,d)\\\pmb{\rm Tr}_s\subset\pmb{\rm Tr}\setminus\{\pmb{1},\pmb{2}\}}}\,\underset{c_i,d_i\in\pmb{t}_i}{\widetilde{\sum}}\,\underset{d_0\in\pmb{2} \setminus e}{\widetilde{\sum}}\,C_{c,d}(\pmb{K}\W\shuffle\pmb{\rm g}\W\shuffle g_0)\,{\rm A}\left(\pmb{\rm Tr}_s\W\shuffle\pmb{\rm g}\W\shuffle g_0,e;\{g_k\}\setminus\pmb{\rm g}\right)\,.~~~\label{expan-h+1g}
\eea
Obviously, the expansion in \eref{expan-h+1g} is equivalent to the general formula \eref{expan-hg}, with the replacement $\{g_k\}\to\{g_k\}\cup g_0$ in \eref{expan-hg}. Consequently, \eref{expan-hg} provides the correct recursive expansion for the most general multi-trace YMS amplitudes.

\subsection{A summary of the construction}

The steps of our construction for multi-trace YMS amplitudes can be summarized as follows.

\begin{itemize}
 \item Step-$1$: Determine the lowest-point one, namely, the $4$-point double-trace amplitude.

 \item Step-$2$: Inverting the leading single-soft theorem for the BAS scalar to generate double-trace amplitudes with more external scalars in one trace of two.

 \item Step-$3$: Derive the double-soft theorems for external scalars belong to a length-$2$ trace, by using the double-trace amplitudes obtained in Step-$2$.

 \item Step-$4$: Inverting the resulted subleading double-soft theorem in Step-$3$, to construct $m$-trace amplitudes with length-$2$ traces.

 \item Step-$5$: Inverting the leading single-soft theorem for the BAS scalar, to construct general $m$-trace amplitudes without external gluon.

 \item Step-$6$: Inverting the subleading single-soft theorem for the gluon to insert external gluons.
\end{itemize}
The whole construction based on the assumption of the universality of soft behaviors. Unlike the construction for single-trace YMS and pure YM amplitudes in section. \ref{sec:BottomUp}, the $4$-point double-trace amplitude, which serve as the starting point of bottom up construction, can not be uniquely determined by imposing general physical requirements such as the correct mass dimension, the Lorentz invariance, the linear dependence of each polarization, and so on. The reason is the freedom for choosing $4$-point interaction. We fixed this starting point by applying the dimensional reduction on $4$-point YM amplitude.

The order of steps listed above is chosen for deriving double-soft theorems for scalars belonging to a length-$2$ trace as simple as possible. Supposing the desired double-soft theorems are given, one can modify the order of steps, or the representation of the $4$-point double-trace amplitude, to get different alternative formulas of multi-trace YMS amplitudes. For example, one can invert the subleading double-soft theorem for scalar to insert length-$2$ traces into the single-trace YMS amplitude in \eref{expan-YMS-recur}, then invert the leading single-soft theorem for scalar to insert external scalars into length-$2$ trace. The resulted formula is found to be
\bea
{\cal A}_{\rm YMS}\left(\pmb{1}|\pmb{2}|\cdots|\pmb{m};\{g_k\}||\pmb\sigma\right)&=&\underset{\pmb{\rm g}\subset\{g_k\}}{\sum}\,\sum_{\substack{\pmb{K}(\pmb{\rm Tr}_s,c,d)\\\pmb{\rm Tr}_s\subset\pmb{\rm Tr}\setminus\pmb{1}}}\,\underset{c_i,d_i\in\pmb{t}_i}{\widetilde{\sum}}\,C'_{c,d}(\pmb{K}\W\shuffle\pmb{\rm g})\,{\rm A}'\left(\pmb{\rm Tr}_s\W\shuffle\pmb{\rm g},p;\{g_k\}\setminus\{\pmb{\rm g}\cup p\}\right)\,,~~~~\label{expan-hg-2}
\eea
where a fiducial gluon $p$ is introduced. This time $\pmb{\rm g}$ is an ordered set whose elements belong to $\{g_k\}\setminus p$. The explicit formulas of $C_{c,d}(\pmb{K}\W\shuffle\pmb{\rm g})$ and ${\rm A}'(\pmb{\rm Tr}_s\W\shuffle\pmb{\rm g},p;\{g_k\}\setminus\pmb{\rm g})$ are given as
\bea
C'_{c,d}(\pmb{K}\W\shuffle\pmb{\rm g})=\epsilon_p\cdot T_{\rho_1}\cdot T_{\rho_2}\cdots T_{\rho_{s+h}}\cdot Y_{\rho_{s+h}}\,,~~~\label{defin-C-3}
\eea
and
\bea
{\rm A}'\left(\pmb{\rm Tr}_s\W\shuffle \pmb{\rm g},p;\{g_k\}\setminus\{\pmb{\rm g}\cup p\}\right)={\cal A}_{\rm YMS}(1,\{2,\cdots,n-1\}\shuffle\{\pmb{K}[\pmb{\rm Tr}_s,c,d]\W\shuffle \pmb{\rm g},p\},n|\pmb{u}_1|\cdots|\pmb{u}_{m-s-1}||\pmb\sigma)\,,~~~\label{defin-A-2}
\eea
which is another type of expansion formula in \cite{Du:2017gnh}. Another example is, one can rewrite the $4$-point double-trace amplitude as
\bea
{\cal A}_{\rm YMS}\left(\pmb{1}|\pmb{2}||\pmb\sigma\right)&=&(k_a\cdot k_1)\,{\cal A}_{\rm BAS}(2,1,a,b||\pmb\sigma)\,,
\eea
then insert more length-$2$ traces, subsequently insert more scalars into each trace, finally insert external gluons. This manipulation yields
\bea
&&{\cal A}_{\rm YMS}\left(\pmb{1}|\pmb{2}|\cdots|\pmb{m};\{g_k\}||\pmb\sigma\right)\nn
&=&\underset{\pmb{\rm g}\subset\{g_k\}}{\sum}\,\sum_{\substack{\pmb{K}(\pmb{\rm Tr}_s,c,d)\\\pmb{\rm Tr}_s\subset\pmb{\rm Tr}\setminus\{\pmb{1},\pmb{2}\}}}\,\underset{c_i,d_i\in\pmb{t}_i}{\widetilde{\sum}}\,\underset{c_0\in\pmb{1} \setminus 2}{\widetilde{\sum}}\,\underset{d_0\in\pmb{2} \setminus b}{\widetilde{\sum}}\,C''_{c,d}(\pmb{K}\W\shuffle\pmb{\rm g})\,{\rm A}''\left(2,\pmb{\rm Tr}_s\W\shuffle\pmb{\rm g},b;\{g_k\}\setminus\pmb{\rm g}\right)\,,~~~~\label{expan-hg-3}
\eea
where
\bea
C''_{c,d}(\pmb{K}\W\shuffle\pmb{\rm g})=k_a\cdot T_{\rho_1}\cdot T_{\rho_2}\cdots T_{\rho_{s+h}}\cdot k_1\,,~~~\label{defin-C-4}
\eea
and
\bea
{\rm A}''\left(2,\pmb{\rm Tr}_s\W\shuffle \pmb{\rm g},b;\{g_k\}\setminus\pmb{\rm g}\right)={\cal A}_{\rm YMS}(2,\mathsf{KK}[\pmb{1},2,c_0],c_0,\pmb{K}[\pmb{\rm Tr}_s,c,d]\W\shuffle \pmb{\rm g},d_0,\mathsf{KK}[\pmb{2},d_0,b],b|\pmb{u}_1|\cdots|\pmb{u}_{m-s-2}||\pmb\sigma)\,.\nn~~~\label{defin-A-3}
\eea
This is form of expansion formula was also proposed in \cite{Du:2017gnh}.  Thus, the construction of amplitudes by soft-behavior provides a really natural way for understanding distinct expansion formulas of a same amplitude.

The expansions in \eref{expan-hg} and \eref{expan-hg-3} are explicitly gauge invariant for each polarization carried by external gluons, since in our method the gauge invariance is automatically manifest when inserting the gluon.

\section{Conclusion and discussion}
\label{sec-conclu}
In this paper, the expansion formula of multi-trace YMS amplitudes were determined by soft behaviors. Instead of supposing the universality and the expansion basis, we took a new bootstrap approach: determining the simplest case which contains two scalar trace and two scalars in each trace, then inserting scalars to a trace such that one trace involves more scalars. Based on these two steps, we get an amplitude with a nontrivial double-soft behavior for scalars in the length-2 trace. When this double-soft factor was obtained, together with known single-soft behaviors of scalars and gluons, it was further used to determine all other multi-trace amplitudes. The construction in this work, together with the well known double copy, in fact also shows that the EYM amplitudes satisfy the same expansion formulas as the YMS amplitudes. This agrees with the known result \cite{Du:2017gnh}.

There are many possible extension discussions. First, it will be interesting to study the relationship between collinear limit and the decomposition formula. Second, the generalization to loop amplitude is another possible direction. Last but not least, since the inverse soft limit method has been used to obtain formulas of helicity amplitudes in four dimensions, we expect that this approach, which does not involve momentum shift, may become another effective tool for studying helicity amplitudes.

\section*{Acknowledgments}
The authors would thank Prof. Song He for helpful comments on the $4$-point interaction in YMS theory. YD is supported by NSFC under Grant No. 11875206. KZ is supported by NSFC under Grant No. 11805163.


\appendix

\section{Absence of pole in coefficients when expanding to BAS basis}
\label{sec-ap}

In this section we give a simple proof for the statement that one can always find a formula, in which coefficients are polynomials of kinematical Lorentz invariants without any pole, when expanding tree amplitudes with massless propagators to BAS KK basis. The proof includes three steps.

\textbf{Step-$1$:}
As well known, any tree amplitudes (kinematic part without coupling constants) with massless propagators can be decomposed into the summation over Feynman diagrams with only cubic interactions as
\bea
{\cal A}=\sum_{\{\Gamma\}}\,{N_\Gamma\over\prod_i\,D_i}\,,~~\label{decom}
\eea
where $\{\Gamma\}$ denotes the set of corresponding diagrams, and the propagators are encoded as $1/D_i$. The key technic is, by inserting $\prod_j{D_j\over D_j}$, one can split higher-point interactions into cubic ones, with additional propagators $1/D_j$, and additional factor $\prod_jD_j$ which can be absorbed into numerators $N_\Gamma$.

\begin{figure}
  \centering
   \includegraphics[width=5cm]{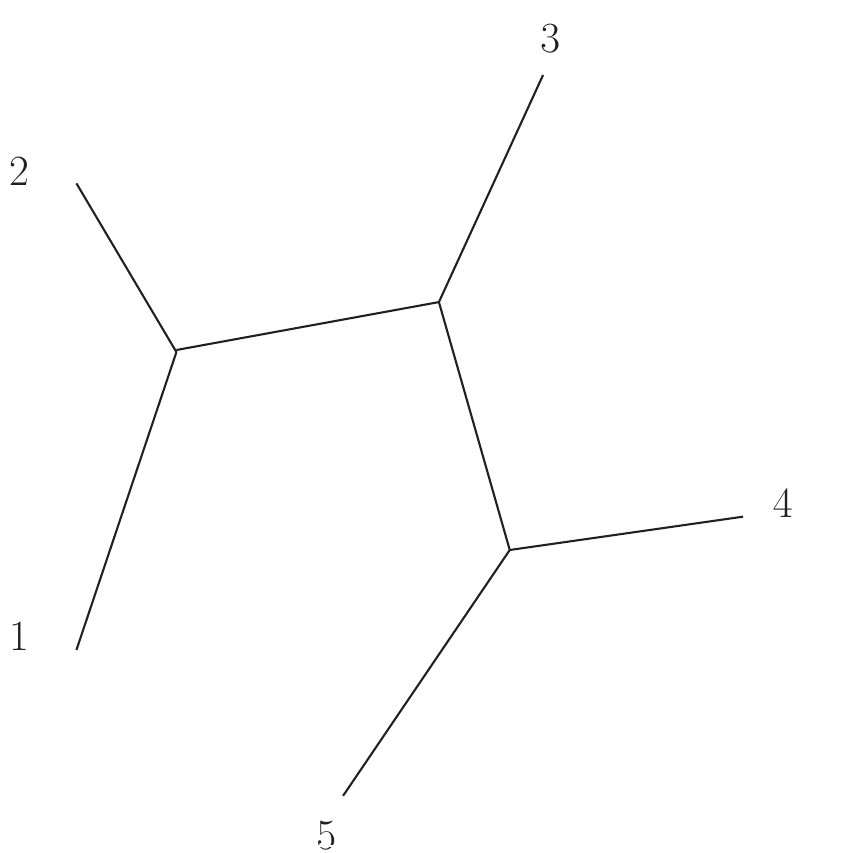}
   \includegraphics[width=5cm]{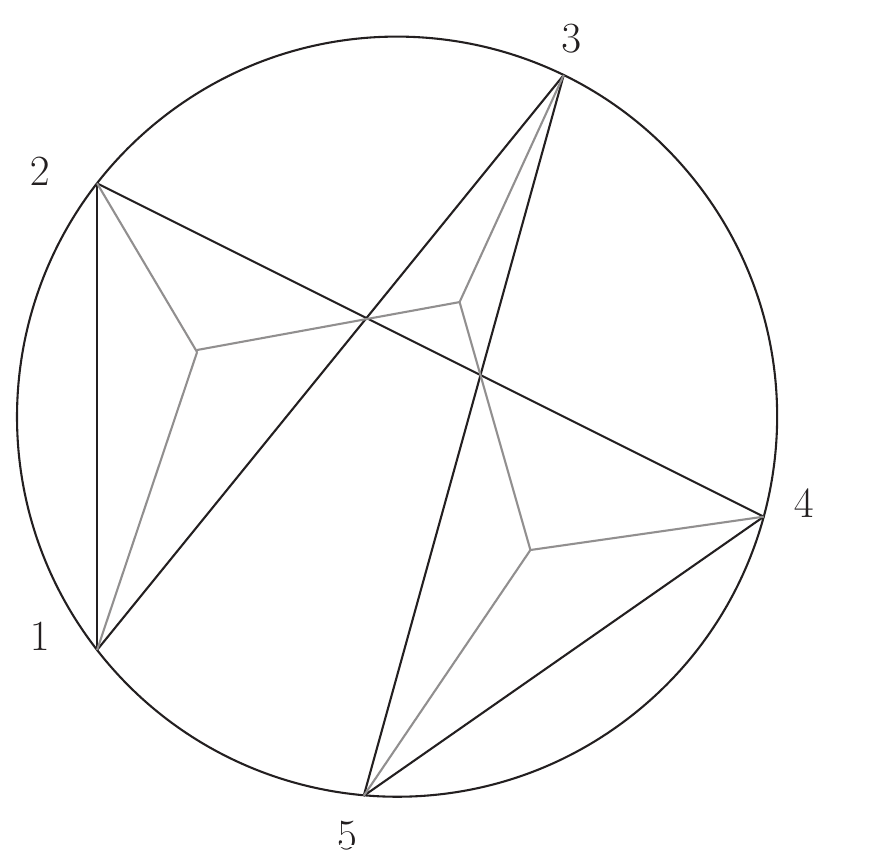} \\
  \caption{Diagrammatical technic for mapping Feynman diagram to BAS amplitude.}\label{example}
\end{figure}

\textbf{Step-$2$:}
In the summation \eref{decom}, each diagram can be mapped to a unique BAS amplitude whose two orderings allow only one diagram, up to an overall $\pm$ sign \cite{Cachazo:2013iea}. For example, using the diagrammatical technic in Figure. \ref{example}, the $5$-point Feynman diagram is mapped to ${\cal A}_{\rm BAS}(1,2,3,4,5||1,2,4,5,3)$. Then the decomposition in \eref{decom} is turned to
\bea
{\cal A}=\sum_{\{\Gamma\}}\,{\rm sign}(\pm)\,N_\Gamma\,{\cal A}_{\rm BAS}(\pmb\sigma||\W{\pmb\sigma})\big|_\Gamma\,,~~\label{decom2}
\eea
where ${\cal A}_{\rm BAS}(\pmb\sigma||\W{\pmb\sigma})\big|_\Gamma$ stands for BAS amplitudes which allow only one diagram belong to $\{\Gamma\}$.

\textbf{Step-$3$:}
The tree BAS amplitudes satisfy the KK relation
\bea
{\cal A}_{\rm BAS}(a,\pmb\a,b,\pmb\b||\pmb\sigma)=(-)^{n_{\pmb\b}}\,{\cal A}_{\rm BAS}(a,\pmb\a\shuffle\pmb\b^T,b||\pmb\sigma)\,,~~\label{KK}
\eea
thus one can use this relation to expand any ${\cal A}_{\rm BAS}(\pmb\sigma||\W{\pmb\sigma})\big|_\Gamma$ into amplitudes ${\cal A}_{\rm BAS}(a,\pmb\sigma',b||\W{\pmb\sigma})$ with $a$ and $b$ are fixed at two ends in the first ordering, and reorganize the summation in \eref{decom2} as
\bea
{\cal A}=\sum_{\pmb\sigma'}\,C(\pmb\sigma',\W{\pmb\sigma})\,{\cal A}_{\rm BAS}(a,\pmb\sigma',b||\W{\pmb\sigma})\,,~~\label{decom3}
\eea
where $\pmb\sigma'$ denotes permutations among legs belong to $\{i\}\setminus\{a,b\}$, with $\{i\}$ the full set of all external legs.
The set of these ${\cal A}_{\rm BAS}(a,\pmb\sigma',b||\W{\pmb\sigma})$ is called the KK basis.
In the expansion \eref{decom3}, all coefficients $C(\pmb\sigma',\W{\pmb\sigma})$ are combinations of proper ${\rm sign}(\pm)N_\Gamma$ in \eref{decom2},
thus are polynomials without any pole.

Notice that for theories under consideration in this paper, coefficients $C(\pmb\sigma',\W{\pmb\sigma})$ in \eref{decom3} are indeed $C(\pmb\sigma')$, i.e., they are independent of $\W{\pmb\sigma}$. In \eref{decom3}, we assumed the dependence on $\W{\pmb\sigma}$ to keep the generality.

\bibliographystyle{JHEP}

\bibliography{reference}

\end{document}